\documentclass[rmp,preprint,preprintnumbers,amsmath,amssymb]{revtex4}

\usepackage{graphicx}
\usepackage{dcolumn}
\usepackage{bm}

\begin{document}
\renewcommand{\thetable}{\arabic{table}}

\title{Induction Level Determines Signature of Gene Expression Noise in Cellular Systems}
\author{Julia Rausenberger}\thanks{To whom the correspondence should be addressed. E-mail: juhe@fdm.uni-freiburg.de}\affiliation{Institute of Physics and Faculty of Biology, University of Freiburg, Germany}
\author{Christian Fleck, Jens Timmer}\affiliation{Institute of Physics, University of Freiburg, Germany}
\author{Markus Kollmann}\affiliation{Institute for Theoretical Biology, Humboldt-University Berlin, Germany}

\begin{abstract}
Noise in gene expression, either due to inherent stochasticity or to varying inter- and intracellular environment, can generate significant cell-to-cell variability of protein levels in clonal populations.
We present a theoretical framework, based on stochastic processes, to quantify the different sources of gene expression noise taking cell division explicitly into account.
Analytical, time-dependent solutions for the noise contributions arising from the major steps involved in protein synthesis are derived.
The analysis shows that the induction level of the activator or transcription factor is crucial for the characteristic signature of the dominant source of gene expression noise and thus bridges the gap between seemingly contradictory experimental results. Furthermore, on the basis of experimentally measured cell distributions, our simulations suggest that transcription factor binding and promoter activation can be modelled independently of each other with sufficient accuracy.
\end{abstract}
\maketitle

\section*{Introduction}
\begin{figure}
\begin{center}
\includegraphics[width=0.35\textwidth]{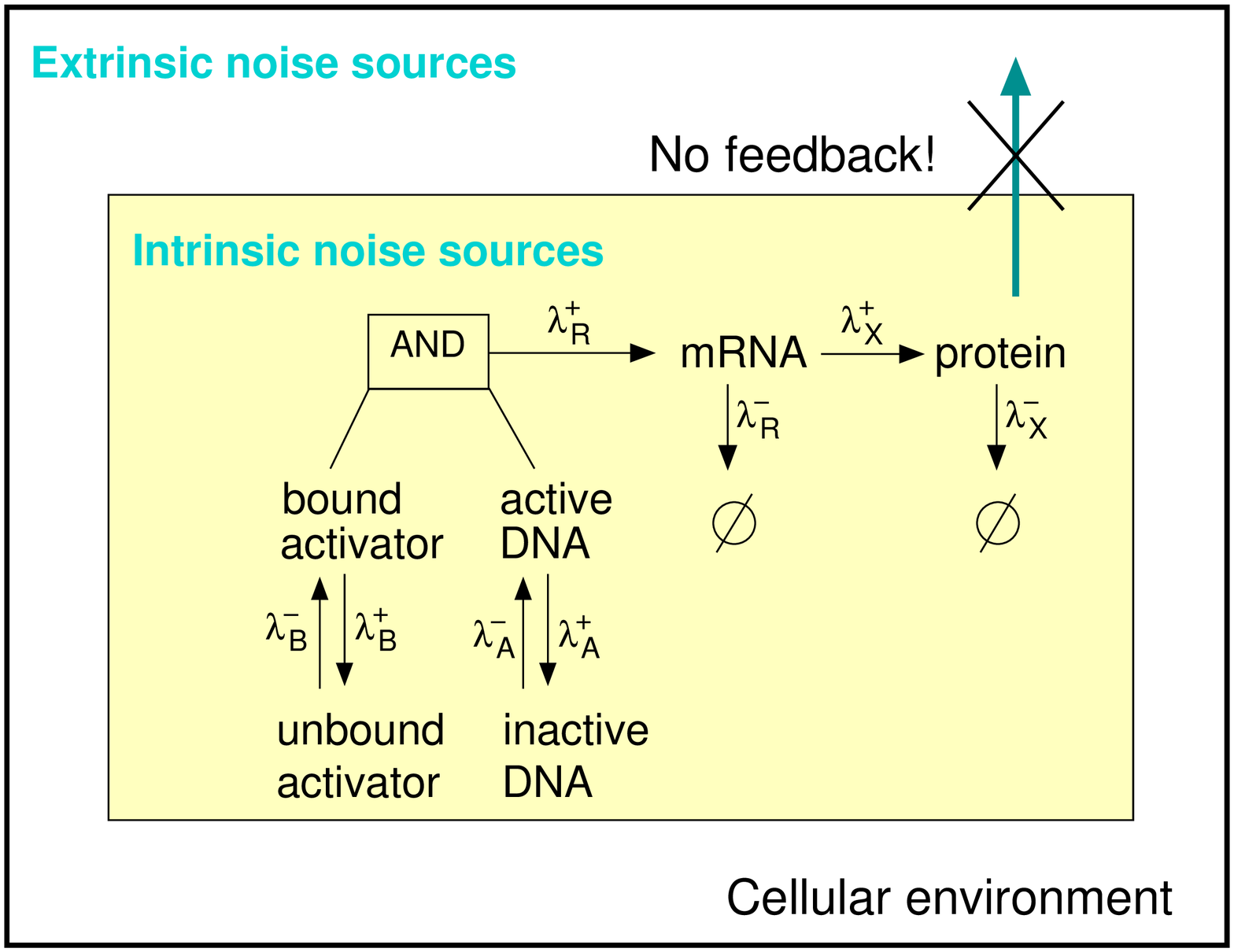}
\caption{\label{fig1} Definition and reaction scheme of single-gene reporter system (yellow box) within intracellular environment (large box). Intrinsic and extrinsic noise can only be distinguished if expression level of reporter system does not influence extrinsic factors. Transition rates are defined in the text.}
\end{center}
\end{figure}

Within a genetically identical population, individual cells show significant phenotypic heterogeneity \cite{Spudich76,RaserOShea05,Avery06}.
This variability directly affects the cell's ability to respond to environmental factors like changes in ligand concentration. 
Especially, reactions underlying protein synthesis are often based on small numbers of molecules, like transcription factors or ribosomes, such that stochastic fluctuations have to be taken into account.\\
A lot of effort has been undertaken to quantify the origins of gene expression noise experimentally and theoretically.
Stochasticity or noise inherent to gene expression seems to be one of the main driving forces for the observed cell-to-cell variability in several experiments which have measured the variance in protein abundances in different cellular systems \cite{Elowitz02,RaserOShea04,Volfson06,BarEven06,Becskei05,Raj06,Blake03,Pedraza05,Golding06,Mettetal06,Newman06,Sigal06,Rosenfeld05,Colman05}. 
Considerable confusion stems from diverging experimental results which have identified different origins for the main contribution to gene expression noise \cite{Kaufmann07} such that a complete picture is still missing. For prokaryotes, translational efficiency was identified as the main source of variability of expression levels consistent with a stochastic model in which proteins are produced in sharp and random bursts \cite{Ozbudak02}. However, later experimental observations in individual living cells either by measuring mRNA levels or by real-time observations at single molecule level indicated that promoter activation predominantly causes gene expression noise \cite{Golding05,Cai06}. Furthermore, extrinsic factors, like the cellular state, were also identified to give the main contribution to phenotypic variations in a clonal population \cite{Rosenfeld05}. 
Similar contradictory results have been found in eukaryotes, where in the budding yeast \textit{Saccharomyces cerevisiae} a two-reporter system, expressing two fluorescent proteins from identical promoters, identified switching between active and inactive promoter states due to slow stochastic chromatin-remodelling events as the by far largest source of noise \cite{RaserOShea04}. 
In later \textit{in-vivo} experiments it was shown for a large set of genes at their native expression levels that the noise has a clear sign of transcriptional origin due to low-copy mRNA molecules \cite{BarEven06,Newman06}. 
Moreover, a direct monitoring of mRNA production from a gene at the resolution of single molecules in mammals revealed strong mRNA bursts dominating gene expression noise \cite{Raj06}. 
In contrast, for human cells, genes at native induction level showed significant noise contribution from long-term variations of the cellular state \cite{Sigal06}. It seems on first sight that no general rule can be given to determine the main sources of gene expression noise. 
Protein levels, however, are often strongly optimized, because they have to allow for precise and reliable information processing within the cell. Any significant deviation from the optimal level would result in reduction of fitness and an evolutionary disadvantage. Thus, random fluctuations are in general detrimental for cellular systems and several regulatory mechanisms have evolved to minimize them. 
Only in rare cases noise can be used to drive phenotypic switching providing a non-genetic mechanism to population heterogeneity, as found for bacterial persistence against antibiotics \cite{Balaban04} and competence for DNA uptake from the environment \cite{Suel06}.\\
In order to track down the individual contributions of the molecular mechanisms involved in protein synthesis several mathematical models have been introduced \cite{Berg78,Thattai01,Swain02,Swain04,Paulsson04,Paulsson05,Friedman06}. 
Some of these models ignore the effect of binomial partitioning by cell division which will lead to strong discrepancies to experiments whenever cellular mRNA is long-lived and appears in low copy number \cite{Paulsson04,Paulsson05,Friedman06}, whereas others lack the dynamic description of mRNA bursts \cite{Berg78,Swain02}.
In the present work we develop an analytical framework which allows for a time-dependent description of gene expression and accounts for effects of symmetric cell division.
We consider a \textit{one-gene-system} consisting of activator/transcription factor (TF) binding (repressor unbinding), promoter activation, transcription, and translation (Fig.~\ref{fig1}).
All gene specific events contribute to the so-called {\it intrinsic noise}. 
Differences between cells, either in global cellular state or in the concentration or activity of any factor that affects gene expression are referred to as {\it extrinsic noise} \cite{Elowitz02}.
Therefore, the cell-to-cell variability of a specific protein in a large clonal population with fixed generation time is characterized by the two contributions of intrinsic and extrinsic noise, summing up to the overall variance of the protein.
Assuming no specific feedback of a produced protein on upstream processes, the intrinsic noise contribution decomposes into partial contributions stemming from activator binding, promoter activation, transcription and translation.
In deriving analytical expressions for these partial contributions to gene expression noise, we discuss limiting cases for mRNA and protein lifetimes.
We show that the magnitudes of the different noise contributions depend strongly on induction level, synthesis rates, and molecule lifetimes associated with each individual gene. 
Therefore, differences in the induction level, e.g., due to different experimental set-ups, might provide a possible explanation for the diverging experimental findings of cell-to-cell variations even in the same organisms.
Furthermore, on the basis of experimentally measured cell distributions of wild-type and over-expressed cells of \textit{E. coli}, our simulations propose that, in prokaryotes, activator binding and promoter activation are independent of each other and thus can be modelled for a good approximation separately.

\section*{Results}
\subsection*{Minimal model for gene expression noise}

Four major steps are involved in a generic model of gene expression in living cells \cite{Paulsson05}: (i) activator/TF binding (repressor unbinding), (ii) promoter (DNA) activation, (iii) transcription, and (iv) translation. 
In the present work we follow a previous approach \cite{Paulsson05} and model synthesis and degradation of mRNA and protein by a {\it birth-and-death} process. 
Activator binding and promoter activation are described as a random telegraph process, because they are assumed to switch randomly between zero and one with exponentially distributed waiting times \cite{Golding05,Raj06}. 
The state of the activator is given by the stochastic variable $B(t)$ switching between $B(t)=1$ and $B(t)=0$, if the activator is bound or unbound, respectively.
Promoter activation can be expected to occur on much slower time scales than activator binding \cite{AlonBuch,Cai06,Golding05,Elf07} such that the time scales can be separated.
Therefore, we approximate activator binding, $B(t)$, as an equilibrated binding process.
The Master equation for the probability to find a promoter in its active, $P(1,t)$, or inactive, $P(0,t)$, state is given by
\begin{eqnarray}
\partial_t P(1,t)&=& -\lambda_A^-P(1,t)+\lambda_A^+P(0,t), \label{mertp} 
\end{eqnarray}
where $\lambda_{A}^+$ represents the switching rate from the inactive to the active state and $\lambda_{A}^-$ the rate for the inverse process.
We define 
\begin{equation}\label{gammaA}
\gamma_A=\lambda_A^++\lambda_A^-
\end{equation}
and denote the initial state of the promoter at time $t_0$ by $\alpha_0$.
The solution for (\ref{mertp}) with initial conditions $P(1,t_0|\alpha_0,t_0)=\delta_{1,\alpha_0}$ reads:
\begin{eqnarray}
P(1,t|\alpha_0,t_0)&=&\frac{\lambda_{A}^+}{\gamma_A}+\left(\delta_{1,\alpha_0}-\frac{\lambda_{A}^+}{\gamma_A}\right)e^{-\gamma_A(t-t_0)}.\label{genactgreensfct}
\end{eqnarray} 
We can assign the stochastic variable $A(t) \in \{0,1\}$ to describe the state of the promoter.
The process of promoter activation seems to have no significant correlations with the cell cycle \cite{Raj06,Golding05}. 
Therefore, we can employ a stationary solution for the auto-covariance function
\begin{eqnarray}
\langle A(t)A(t')\rangle-\langle A(t)\rangle\langle A(t')\rangle
&=&var(A)e^{-\gamma_A|t-t'|}\label{autocorrA}
\end{eqnarray}
with $var(A)=\lambda_A^+\lambda_A^-/\gamma_A^2$, and $\langle A \rangle=\lambda_A^+/\gamma_A$. The mean $\langle A\rangle$ can be interpreted as a measure of the fraction a promoter spends in its active state for which holds $0\leq\langle A\rangle\leq1$ .\\
Transcription and translation are modelled as a \textit{birth-and-death} process.
The corresponding mRNA and protein trajectories are denoted by $R(t)$ and $X(t)$, respectively.
For the conditional probability $P(n_i,t|n_i',t_0)$ to observe $n_i$ molecules ($n_1=R, n_2=X$) in a given cell at time $t$ we obtain the stochastic Master equation
\begin{eqnarray}
\partial_t P(n_i,t|n_i',t_0)&=&\lambda_i^+(t)P(n_i-1,t|n_i',t_0)\nonumber\\
&+& (n_i+1)\lambda_i^-P(n_i+1,t|n_i',t_0)\nonumber\\
&-& (\lambda_i^-n_i+\lambda_i^+(t))P(n_i,t|n_i',t_0)\label{mastertransc}
\end{eqnarray}
where $n_i'$ molecules are observed initially at time $t_0$.
Synthesis rates are given by $\lambda_1^+(t)=\lambda_R^+A(t)B(t)$  for the mRNA or $\lambda_2^+(t)=\lambda_X^+R(t)$ for the proteins.
The corresponding degradation rates, $\lambda_1^-=\lambda_R^-$ and $\lambda_2^-=\lambda_X^-$, are assumed to be independent of molecule number and constant in time. 
Furthermore, the trajectories $A(t)$ and $B(t)$ are assumed to be independent of each other which will be justified below. 
Transcription at time $t$ is only possible if the activator is bound \textit{and} the promoter is in the active state, $A(t)B(t)=1$, cf. Fig.~\ref{fig1}. 
We assume that the translation rate of specific proteins depends linearly on the actual amount of the corresponding mRNA, $R(t)$.

A lot of effort has been undertaken to generate time resolved expression data of single cells \cite{Sigal06}. 
The mathematical description of gene expression becomes more complicated if cell division is included, because cell division introduces another important time scale into the system, the generation time $T_G$.
Previous theoretical approaches avoided this problem by assuming an increased protein degradation rate instead of explicitly taking cell division into account \cite{Paulsson04}. Such an approximation might be justified for systems where effects of binomial partitioning can be neglected and cellular growth and protein synthesis scale linearly in time. In this case the concentrations - but not the absolute copy number of molecules - are approximately constant over cell cycles. However, one should note that experiments measure the total molecule number instead of concentrations.
Measurement of concentrations would require precise knowledge of the individual cell volume at any time.
Recent experiments show that the main effect of cell division seems to be binomial partitioning of molecules \cite{Golding05,Rosenfeld05} such that the assumption of an increased protein degradation does not hold.

If we assume symmetric cell division at time $t_0$ with $t'<t_0<t$ and a binomial distribution of the molecules, the following relationship for the conditional probability $P(n_i,t|n_i',t')$ holds:
\begin{eqnarray}
P(n_i,t|n_i',t')&=&\sum_{n_i^-=0}^{\infty}\sum_{n_i^+=0}^{n_i^-} P(n_i,t|n_i^+,t_0){n_i^-\choose n_i^+}2^{-n_i^-}P(n_i^-,t_0|n_i',t').\label{a4}
\end{eqnarray}
The amounts of molecules before and after the last cell division occurring at time $t_0$ are represented by $n_i^-$ and $n_i^+$, respectively.
To solve Eqn.~(\ref{mastertransc}) we use the generating function $G(s,t|n_i',t')=\sum_{n_i=0}^{\infty}s^{n_i}P(n_i,t|n_i',t')$ and assume a finite number of cell divisions between $t'$ and $t$. After some algebra we arrive (for derivation see SI) at:
\begin{eqnarray} \label{completegenfunct}
&&G(s,t|n_i',t')=\left[2^{-D(t,t')}(s-1)e^{-\lambda_i^-(t-t')}+1\right]^{n_i'}\\ 
&&\qquad\qquad\qquad\times\exp\left[(s-1)\int_{t'}^t \lambda_i^+(t'')e^{-\lambda_i^-(t-t'')}2^{-D(t,t'')}dt''\right],\nonumber
\end{eqnarray}
where $D(t,t')$ is the number of cell divisions that have occurred between two time points $t>t'$.
The mean amount of molecules at time $t>t'$ can be calculated via $\partial_sG|_{s=1}$. 
If we assume that initially the process is Poissonian-distributed the mean amount of molecules is given by
\begin{eqnarray}
\langle n_i(t) \rangle_{i} &=&\langle n_i(t') \rangle_{i} 2^{-D(t',t)}e^{-\lambda_i^-(t-t')}+\int_{t'}^t\lambda_i^+(t'')e^{-\lambda_i^-(t-t'')}2^{-D(t,t'')}dt''\label{mean_n}
\end{eqnarray}
where the average over $R$ or $X$ is denoted by $\langle\, .\, \rangle_{i}$. 
We define the generating function $\Theta(s,t|n_i',t_0)=\sum_{n_i=0}^{\infty}s^{n_i}P(n_i,t|n_i',t_0)$ and use Eqn.~(\ref{completegenfunct}) to obtain the auto-correlation function (\cite{vanKampen92}) via $\partial_s\Theta|_{s=1}$ (for detailed derivation see SI):
\begin{eqnarray}
\langle n_i(t),n_i(t')\rangle_{i} = \langle n_i(t')\rangle_{i}2^{-D(t,t')}e^{-\lambda_i^-(t-t')}. \label{autocorrfunc_n}
\end{eqnarray}
Eqn.~(\ref{autocorrfunc_n}) shows that each cell division halves the magnitude of the auto-correlation function.  
Replacing $n_i$ by $R$ or $X$, we obtain the mean amount and the auto-correlation function of the mRNA or the protein from Eqns.~(\ref{mean_n}) and (\ref{autocorrfunc_n}), respectively.

\subsection*{Partial contributions to gene expression noise}
The heterogeneity in gene expression of a population can be quantified using the standard deviation ($\sigma$) divided by the mean ($\mu$), i.e., $\eta=\sigma/\mu$. The quantity $\eta$ is commonly denoted as noise and provides a physiologically relevant measure of gene expression variability as it quantifies relative fluctuations independent of the expression level.
Two main contributions to the overall variance, $\sigma_{tot}^2$, determine the cell-to-cell variability in the amount of a protein, $X$: the intrinsic variance, $\sigma_{I}^2$, which is distinctive for each gene in its genomic context and an extrinsic noise contribution stemming from the variance $\sigma_{E}^2$, which is due to fluctuations in the intra-cellular environment. If we do not assume any significant feedback of the expressed protein on extrinsic factors (cf.~Fig.~\ref{fig1} and \cite{tenWolde06}), the overall variance in a large clonal population of cells with fixed generation time $T_G$ sums up to:
 \begin{eqnarray}
\sigma_{tot}^2 =\sigma_{I}^2 +\sigma_{E}^2.
\label{a1}
\end{eqnarray}
The extrinsic contributions are usually separated into fluctuations of upstream factors that drive expression directly, e.g.,~a given activator concentration and the cellular state that influences gene expression, e.g.,~via variations in polymerase, ribosome, and protease concentrations.
The average over all possible trajectories of protein copy number $X(t)$, mRNA copy number $R(t)$, promoter activation $A(t)$, and activator binding $B(t)$ is defined by $\langle\, .\, \rangle_{\bf I}=\int\, .\,P[X(t);R(t);A(t);B(t)]dX(t)dR(t)dA(t)dB(t)$ with the multi-index ${\bf I}=(X,R,A,B)$, using e.g.,~$\langle\, .\, \rangle_X=\int\, .\, P[X(t)|R(t)]dX(t)$. Note that $P[X;R;A;B]$ can be written as $P[X|R]P[R|A,B]P[A]P[B]$. 
We can identify different contributions to the {\it intrinsic noise} of protein synthesis:
\begin{eqnarray}
\sigma_{I}^2&=&\left\langle \left(X(t)-\langle X(t)\rangle_{X,R,A,B}\right)^2\right\rangle_{X,R,A,B}\nonumber\\
&=&\left\langle \left(X(t)-\langle X(t)\rangle_X\right)^2\right\rangle_{X,R,A,B} \nonumber\\
&+& \left\langle \left(\langle X(t)\rangle_X-\langle X(t)\rangle_{X,R}\right)^2\right\rangle_{R,A,B} \nonumber\\
&+& \left\langle \left(\langle X(t)\rangle_{X,R}-\langle X(t)\rangle_{X,R,A}\right)^2\right\rangle_{A,B} \nonumber\\
&+& \left\langle \left(\langle X(t)\rangle_{X,R,A}-\langle X(t)\rangle_{X,R,A,B}\right)^2\right\rangle_{B} \label{a3}\\
&=&\sigma^2_B+\sigma^2_A+\sigma^2_R+\sigma^2_X,
\label{a2}
\end{eqnarray}
where the right-hand-side denotes the sum over variances (for derivation see SI) corresponding to the processes of activator binding ($\sigma_B^2$), promoter activation ($\sigma_A^2$), transcription ($\sigma_R^2$), and translation ($\sigma_X^2$), respectively. 
The summing up of the individual variances is only possible if there are not any feedbacks from downstream to upstream processes.
We find for the noise contribution due to translation: $\eta_X^2=\sigma^2_X/\langle X \rangle_{\bf I}^2= 1/\langle X \rangle_{\bf I}$. Also, the binomial distribution of the proteins caused by cell division converges quite rapidly to a Gaussian distribution for an increased amount of molecules. The amount of protein synthesized per mRNA can be estimated to be of the order $10-10^3$ for most systems. Thus, for the experiments considered in this work, we can neglect translational noise in comparison to other noise contributions.\\

In order to make the analytical calculations feasible, we assume a fixed induction level of the activator, $B(t)=B_{eq}$.
Due to the separation of the time scales (fast activator binding and very slow promoter activation), fluctuations from activator binding can be neglected compared to fluctuations from promoter activation and transcription such that the averaging over $B$ cancels out in Eqns.~(\ref{a3}).
Therefore, noise contribution from activator binding, $\sigma_B^2$, is not present and we set ${\bf I}=(X,R,A)$.
Explicit expressions for the different intrinsic noise contributions of promoter activation, $\sigma_A^2$, and transcription, $\sigma_R^2$, can be obtained from Eqns.~(\ref{autocorrA}),(\ref{mean_n}) and (\ref{autocorrfunc_n}) (see SI for a detailed derivation). 
We introduce the abbreviations 
\begin{equation}
{\cal A}=\langle A \rangle B_{eq} \lambda_R^+\lambda_X^+
\end{equation}
representing the average acceleration of protein synthesis in absence of any degradation, and
\begin{equation}
Z_R(T_G)=\frac{1-\exp[-\lambda_R^-T_G]}{1-2^{-1}\exp[-\lambda_R^-T_G]}.
\end{equation}
Assuming infinitely many cell divisions and averaging over all intrinsic processes yields for the mean amount of protein and mRNA:
\begin{eqnarray}
\hspace*{-7mm}&&\langle X(t-t_0)\rangle_{\bf I}=\frac{{\cal A}}{\lambda_R^-} \left[\frac{1}{\lambda_X^-}\left(1-e^{-\lambda_X^-(t-t_0)}\right)-\frac{1-2^{-1}Z_R}{\lambda_R^--\lambda_X^-}\left(e^{-\lambda_X^-(t-t_0)}-e^{-\lambda_R^-(t-t_0)} \right)  \right]\nonumber\\
\hspace*{-7mm}&&\quad+ \frac{2^{-1}}{1-2^{-1}e^{-\lambda_X^-T_G}}e^{-\lambda_X^-(t-t_0)}\frac{{\cal A}}{\lambda_R^-} \left[\frac{1}{\lambda_X^-}\left(1-e^{-\lambda_X^-T_G}\right) -\frac{1-2^{-1}Z_R}{\lambda_R^--\lambda_X^-}\left(e^{-\lambda_X^-T_G}-e^{-\lambda_R^-T_G} \right)  \right]\label{averagedmeanx}\\
\hspace*{-7mm}&&\nonumber\\
\hspace*{-7mm}&&\langle R(t-t_0) \rangle_{\bf I}= \langle A \rangle B_{eq} \frac{\lambda_R^+}{\lambda_R^-}\left[1-(1-2^{-1}Z_R)e^{-\lambda_R^-(t-t_0)}\right] . \label{averagedmeanr}
\end{eqnarray}
The terms involving the generation time, $T_G$, reflect the memory of the contributions that have been generated in previous generations and have been passed over to the actual generation.
Since we account for cell division, the results can be directly compared to time-resolved expression data of protein levels.\\
In the following we discuss three important physiological limiting cases in detail. The derived asymptotic expressions for the mean and the variances are valid for any time $t$ within a given cell cycle, $0\leq t-t_0 \leq T_G$, where $t_0$ denotes the time point of the last cell division. In the first two cases we omit the explicit formulas for the corresponding noise contributions $\eta=\sigma/\mu$, since these formulas are easily derived but do not contain new information with respect to the derived mean and variances.\\ 

\paragraph*{Case I: Short mRNA lifetimes and long protein lifetimes, $(\lambda_R^-)^{-1}\ll T_G\ll(\lambda_X^-)^{-1}$.}
This is the most likely physiological case.
We obtain for $t-t_0,T_G\gg (\lambda_R^-)^{-1}$ the following asymptotic expressions for the average amount of protein and mRNA:
\begin{eqnarray}
\langle X(t-t_0) \rangle_{\bf I} &=&\frac{{\cal A}}{\lambda_R^-}\left[(t-t_0)+T_G\right] \label{meanXcase1}\\
\langle R(t-t_0) \rangle_{\bf I} &=&\langle A \rangle B_{eq} \frac{\lambda_R^+}{\lambda_R^-}. 
\end{eqnarray}
Note that Eqn.~(\ref{meanXcase1}) implies a linear increase of the mean amount of protein in time as well as a doubling of protein synthesized over one cell cycle, $\langle X(T_G)\rangle_{\bf I}=2\langle X(t_0)\rangle_{\bf I}$. 
The stationary mRNA level is recovered immediately after cell division.
The noise contributions from transcription and promoter activation are also time-dependent (for derivation see SI) and read in the limit $t-t_0,T_G\gg (\lambda_R^-)^{-1}, \gamma_A^{-1}$:
\begin{eqnarray}
\sigma_R^2(t-t_0)&=&2{\cal A}\frac{\lambda_X^+}{(\lambda_R^-)^2}\left[t-t_0+\frac{1}{3}T_G\right]\label{case1Trans}\\
\sigma_A^2(t-t_0)&=&2 {\cal A}^2 \frac{var(A)}{\langle A \rangle^2}\frac{1}{(\lambda_R^-)^2\gamma_A}\left[(t-t_0)+\frac{1}{3}T_G\right].\label{case1Prom}
\end{eqnarray}

\paragraph*{Case II: Long mRNA lifetimes and long protein lifetimes, $(\lambda_R^-)^{-1},(\lambda_X^-)^{-1}\gg T_G$.}
Molecule lifetimes are significantly larger than the generation time and the switching rate of promoter activation, which result in the asymptotic expressions for the means:
\begin{eqnarray}
\langle X(t-t_0) \rangle_{\bf I} &=& \frac{1}{2}{\cal A}\left[(t-t_0)^2+T_G^2\right] \\
\langle R(t-t_0) \rangle_{\bf I} &=& \langle A \rangle B_{eq} \lambda_R^+\left[t-t_0+T_G\right]. 
\end{eqnarray}
The mean amount of mRNA increases linearly whereas the mean amount of protein increases quadratically within one cell cycle. Both the mRNA and the protein synthesized over one cell cycle are doubled.
In this asymptotic case, dilution due to cell division is the only mechanism which determines the correlation times. 
Regarding transcriptional noise, we find in the limit $t-t_0,T_G \gg \gamma_A^{-1}$ strong contributions from previous generations, reflected by several terms involving $T_G$:
\begin{equation}
\sigma_R^2(t-t_0)=\frac{1}{3}{\cal A}\lambda_X^+\left[(t-t_0)^3 + T_G^2 (t-t_0) + \frac{2}{3}T_G^3\right].\label{case2Trans} 
\end{equation}
Noise due to promoter activation in the limit $t-t_0,T_G \gg \gamma_A^{-1}$ takes the functional form:
\begin{eqnarray}
\sigma_A^2(t-t_0)&=&\frac{2}{3}{\cal A}^2 \frac{var(A)}{\langle A \rangle^2}\frac{1}{\gamma_A}\left[(t-t_0)^3 + T_G(t-t_0)^2 \index{}+12 T_G^2(t-t_0) +14 T_G^3\right].\label{case2Prom}
\end{eqnarray}
Experimentally one could test this case using a set-up with artificially stabilized mRNA.

\paragraph*{Case III: Short mRNA lifetimes and short protein lifetimes, $(\lambda_R^-)^{-1},(\lambda_X^-)^{-1}\ll T_G$.}
In the exceptional case that both protein and mRNA lifetimes are significantly shorter than the generation time, memory over generations is eliminated and the stationary protein level is recovered immediately after cell division. 
The solutions are consequently time-independent and agree with those found earlier by Paulsson \cite{Paulsson04,Paulsson05} for the mean amount of protein and mRNA:
 \begin{eqnarray}
\langle X \rangle_{\bf I} &=& \frac{{\cal A}}{\lambda_R^-\lambda_X^-}\\  
\langle R \rangle_{\bf I} &=& \langle A \rangle  B_{eq}\frac{\lambda_R^+}{\lambda_R^-}
\end{eqnarray}
and the variances
\begin{eqnarray}
\sigma_R^2&=&{\cal A} \frac{\lambda_X^+}{\lambda_X^-\lambda_R^-(\lambda_R^-+\lambda_X^-)} \label{case3Trans}\\
\sigma_A^2&=&\frac{{\cal A}^2 var(A)}{\langle A\rangle^2\lambda_R^-\lambda_X^-(\gamma_A+\lambda_R^-)(\lambda_R^- +\lambda_X^-)}\left(1+\frac{\lambda_R^-}{\gamma_A+\lambda_X^-}\right).\label{case3Prom}
\end{eqnarray}

\noindent The noise contributions from transcription and promoter activation to gene expression noise are given by 
\begin{eqnarray}
\eta_R^2\hspace*{-0.05cm}&=&\hspace*{-0.05cm}\frac{1}{\langle R \rangle_{\bf I}}\frac{\lambda_X^-}{\lambda_R^-+\lambda_X^-}\\
\eta_A^2\hspace*{-0.05cm}&=&\hspace*{-0.05cm}\frac{var(A)}{ \langle A \rangle^2 }\frac{\lambda_R^-}{\gamma_A+\lambda_R^-}\frac{\lambda_X^-}{\lambda_R^-+\lambda_X^-} \hspace*{-0.1cm}\left( 1+\frac{\lambda_R^-}{\gamma_A+\lambda_X^-} \right)
\end{eqnarray}
and have been calculated by Paulsson as the normalized stationary variance with the same result (\cite{Paulsson05}, Eqn.~(4)).

\subsection*{Noise regimes account for different experimental observations}

\begin{figure}
\begin{center}
\includegraphics[width=.3\textwidth]{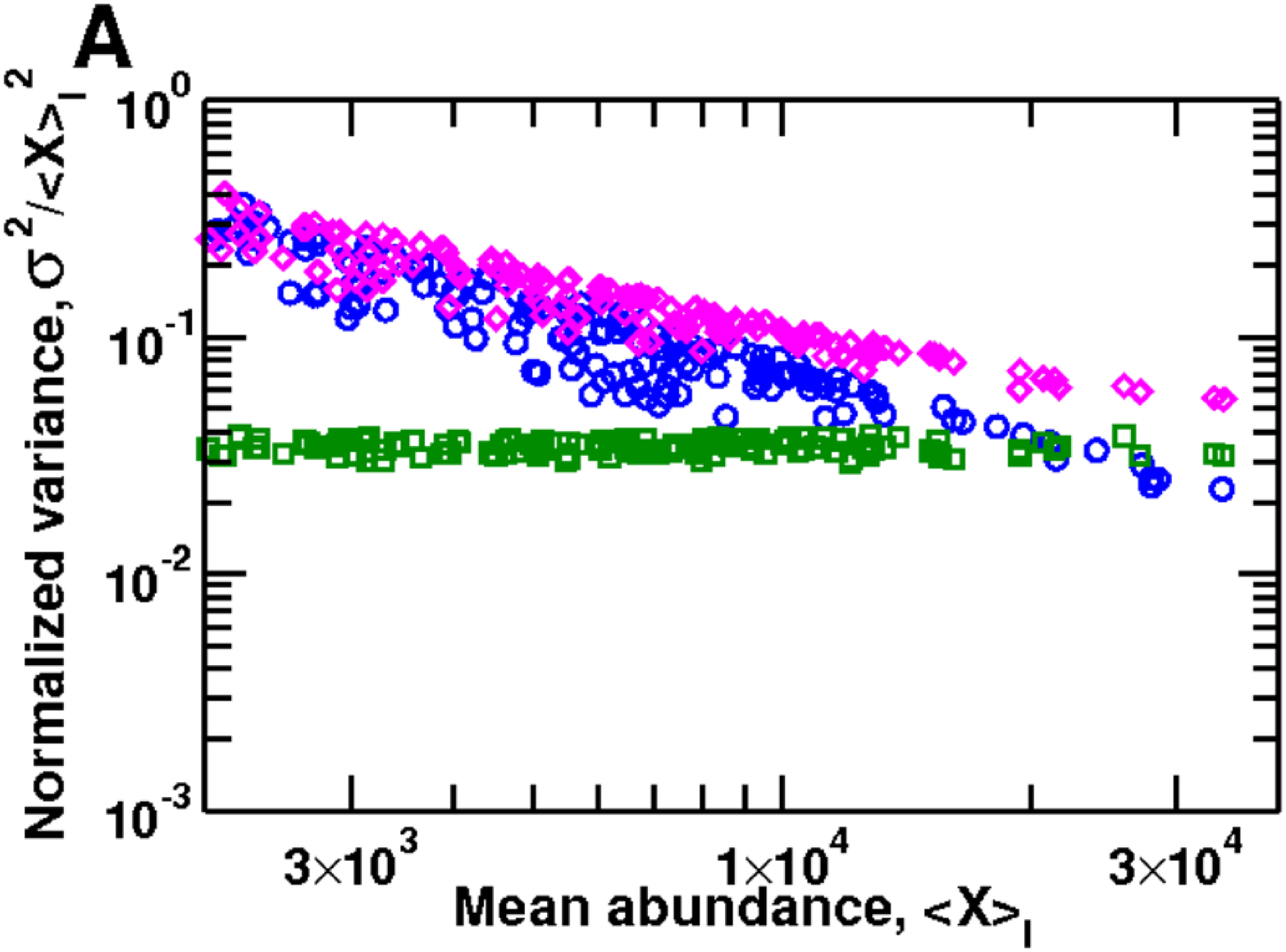}\\
\vspace*{3mm}
\includegraphics[width=.3\textwidth]{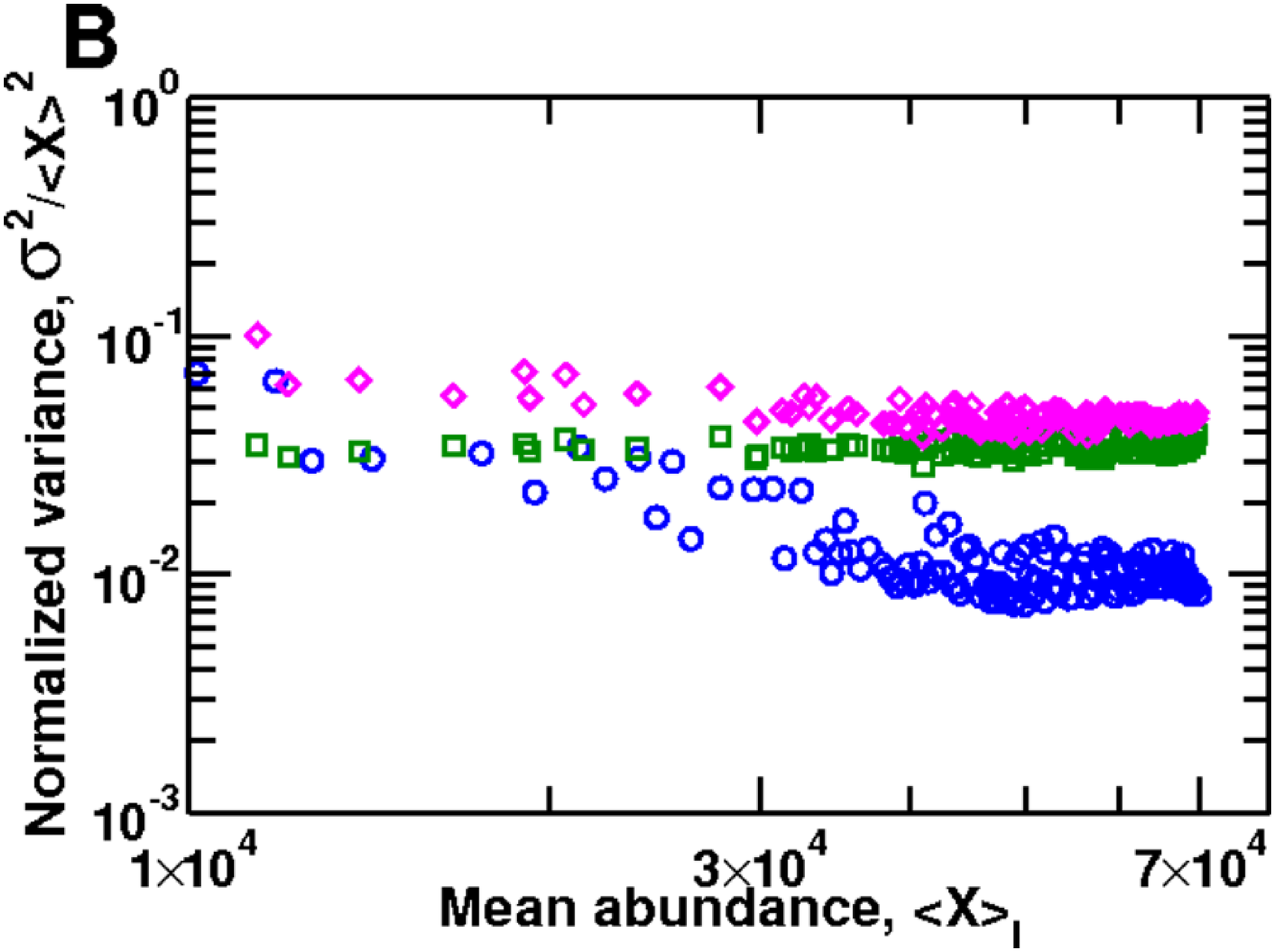}
\caption{\label{fig2}Simulated mean abundance vs. noise for different genes. Transcriptional contribution (blue circles), noise from promoter activation (green squares) and overall noise (magenta diamonds).\\ 
\textbf{A)} Low induced activator, $B_{eq}=0.07$, leads to $\sigma_A^2/\sigma_R^2<1$ such that transcriptional noise dominates. \textbf{B)} Highly induced activator, $B_{eq}=0.7$, leads to $\sigma_A^2/\sigma_R^2>1$ such that noise from promoter activation dominates.}
\end{center}
\end{figure}

Different origins of noise have been proposed and measured by several experimental groups \cite{RaserOShea04,BarEven06,Raj06,Newman06,Sigal06,Rosenfeld05,Ozbudak02,Golding05,Cai06}. Recently, Kaufmann and van Oudenaarden critically reviewed these experimental observations \cite{Kaufmann07}. The contradictory results, even in the same eukaryotic organism \textit{S. cerevisiae}, support the idea that gene expression is influenced by more than one main driving source.
In the following we focus on the budding yeast \textit{S. cerevisiae} and the contradictory experimental results found by Bar-Even et al., Newman et al. and Raser and O'Shea  \cite{RaserOShea04,BarEven06,Newman06}. 
Bar-Even et al. as well as Raser and O'Shea developed the same mathematical model to describe gene expression: it contains the processes of gene/promoter activation, transcription and translation (see Supporting Online Materials of \cite{RaserOShea04} and \cite{BarEven06}). However, the authors interpreted the theoretical  results according to their experimental observations.
Raser and O'Shea \cite{RaserOShea04} measured the intrinsic noise strength of the \textit{PHO5} and \textit{PHO84} promoters at different rates of gene expression in promoter constructs. 
They concluded that the noise intrinsic to gene expression is promoter-specific: noise generation at the \textit{PHO5} promoter depends on stochastic promoter activation due to chromatin remodelling. 
In contrast, Bar-Even et al. \cite{BarEven06} investigated native expression of 43 genes in each of 11 conditions, whereas Newman et al. \cite{Newman06} presented an extensive overview of protein noise for more than 2500 proteins expressed from their endogenous promoter and natural chromosomal position by the use of a combination of high-throughput flow cytometry and a library of GFP-tagged yeast strains. Both latter studies concluded that random \textit{birth-and-death} of low-copy mRNA molecules describe the large observed variations quite well: for the great majority of proteins the noise level is inversely proportional to the mean protein abundance implying a clear signature of a Poisson process.
The obvious question arises: what is the predominant source of noise: promoter activation, as suggested by Raser and O'Shea, or mRNA fluctuations due to low copy number as proposed by Bar-Even et al. and Newman et al.?
An explanation for these contradictory experimental results is given by our stochastic model considering activator binding explicitly.
For approximately constant activator concentration and high amount of protein synthesized per mRNA, the intrinsic variance $\sigma_I^2$ of Eqn.~(\ref{a2}) reduces to $\sigma_I^2=\sigma_A^2+\sigma_R^2$.
Hence the ratio $\sigma_A^2/\sigma_R^2$ determines the predominant source of noise: if $\sigma_A^2/\sigma_R^2\gg 1$, promoter activation will be the dominant process while in the case of $\sigma_A^2/\sigma_R^2\ll 1$ the major part of gene expression noise is due to transcription.
For long protein lifetimes and short mRNA lifetimes (Eqns.~(\ref{case1Trans}), (\ref{case1Prom})) the ratio is given by
\begin{equation}\label{sigmaARratio}
\frac{\sigma_A^2}{\sigma_R^2}=B_{eq} \frac{var(A)}{\langle A \rangle} \frac{\lambda_R^+}{\gamma_A}=B_{eq}\frac{\lambda_A^-\lambda_R^+}{\gamma_A^2}.
\end{equation} 
Note that this ratio is time- and cell cycle-independent although $\sigma_R$, Eqn.~(\ref{case1Trans}), and $\sigma_A$, Eqn.~(\ref{case1Prom}), both depend on the cell cycle time, $t-t_0$, and generation time, $T_G$.
For short mRNA and short protein lifetimes we obtain from Eqns.~(\ref{case3Trans}), (\ref{case3Prom}) the following ratio:
\begin{equation}\label{sigmaARratiocase3}
\frac{\sigma_A^2}{\sigma_R^2}=B_{eq} \frac{var(A)}{\langle A \rangle} \frac{\lambda_R^+}{\gamma_A+\lambda_R^-}\left(1+\frac{\lambda_R^-}{\gamma_A+\lambda_X^-}\right).
\end{equation} 
If we subsequently assume, as it has been done by Bar-Even et al. and Raser and O'Shea, that protein lifetimes are longer, such that switching between promoter states occurs more frequently than a protein degradation event, $\lambda_R^-\gg\gamma_A\gg\lambda_X^-$, Eqn.~(\ref{sigmaARratiocase3}) reduces to Eqn.~(\ref{sigmaARratio}). 
It follows from Eqns.~(\ref{sigmaARratio}) and (\ref{sigmaARratiocase3}) that for fixed rates the probability of activator binding, $B_{eq}$, determines the value of the ratio $\sigma_A^2/\sigma_R^2$, i.e. the induction level of the activator selects the predominant source of noise.
Thus, we expect for highly expressed genes, $B_{eq}\to 1$, to show signature of noise from promoter activation provided $\lambda_A^- \lambda_R^+>\gamma_A^2$ 
since in this case we find $\sigma_A^2/\sigma_R^2> 1$.
In contrast, we expect for low induced genes, $B_{eq}\ll 1$, to show signature of Poissonian noise from mRNA synthesis, since in this case $\sigma_A^2/\sigma_R^2<1$ holds given $\lambda_A^-\lambda_R^+/\gamma_A^2$ is not too large.
Therefore, the induction level of the activator, $B_{eq}$, provides an excellent explanation for the observation of different noise contributions even in the same organism. 
In fact, activator induction is expected to be quite low for experiments with native genes \cite{Newman06,Raj06}. Bar-Even et al. investigated native genes implying a large set of low induced genes, $B_{eq}\ll 1$, such that transcription is the prevailing source of noise. 
In contrast, Raser and O'Shea constructed yeast strains that expressed CFP and GFP proteins from identical promoters. In constructs activator induction is very high, $B_{eq}\rightarrow 1$, such that we expect that promoter activation noise is the dominant noise contribution.
In Fig.~\ref{fig2} we present the mean protein abundance vs. noise. 
In order to select arbitrary time points $t$ within a given cell cycle, $0\leq t-t_0\leq T_G$, we use the full expressions for the mean amount of protein, Eqn.~(\ref{averagedmeanx}), and noise contributions (see SI), because the approximations presented in Eqns.~\eqref{meanXcase1}, \eqref{case1Trans} and \eqref{case1Prom} are only valid in the asymptotic limit $t-t_0\gg(\lambda_R^-)^{-1}$. 
We calculate the mean amount of protein and the noise contributions for several genes at randomly selected time points for several induction levels of the activator binding $B_{eq}$. 
In Fig.~\ref{fig2}A we assume a low induction of the activator, where the mean induction level $B_{eq}$ equals 0.07, and therefore mimics the experimental set-up of Bar-Even et al.
The noise contribution arising from transcription (blue circles) dominates the overall noise (magenta diamonds).
In Fig.~\ref{fig2}B, however, noise from promoter activation (green squares) overrules noise from transcription. This can be arranged with an highly induced activator, with mean induction level $B_{eq}=0.7$, which mimics the experimental set-up of Raser and O'Shea. 
We conclude that both experimental scenarios can be qualitatively reproduced very well with our stochastic model by varying the induction level of the activator binding.

\subsection*{Activator binding and promoter activation determine population distribution}
\begin{figure}
\begin{center}
\includegraphics[width=.3\textwidth]{Fig3A.eps}\\
\vspace*{3mm}
\includegraphics[width=.3\textwidth]{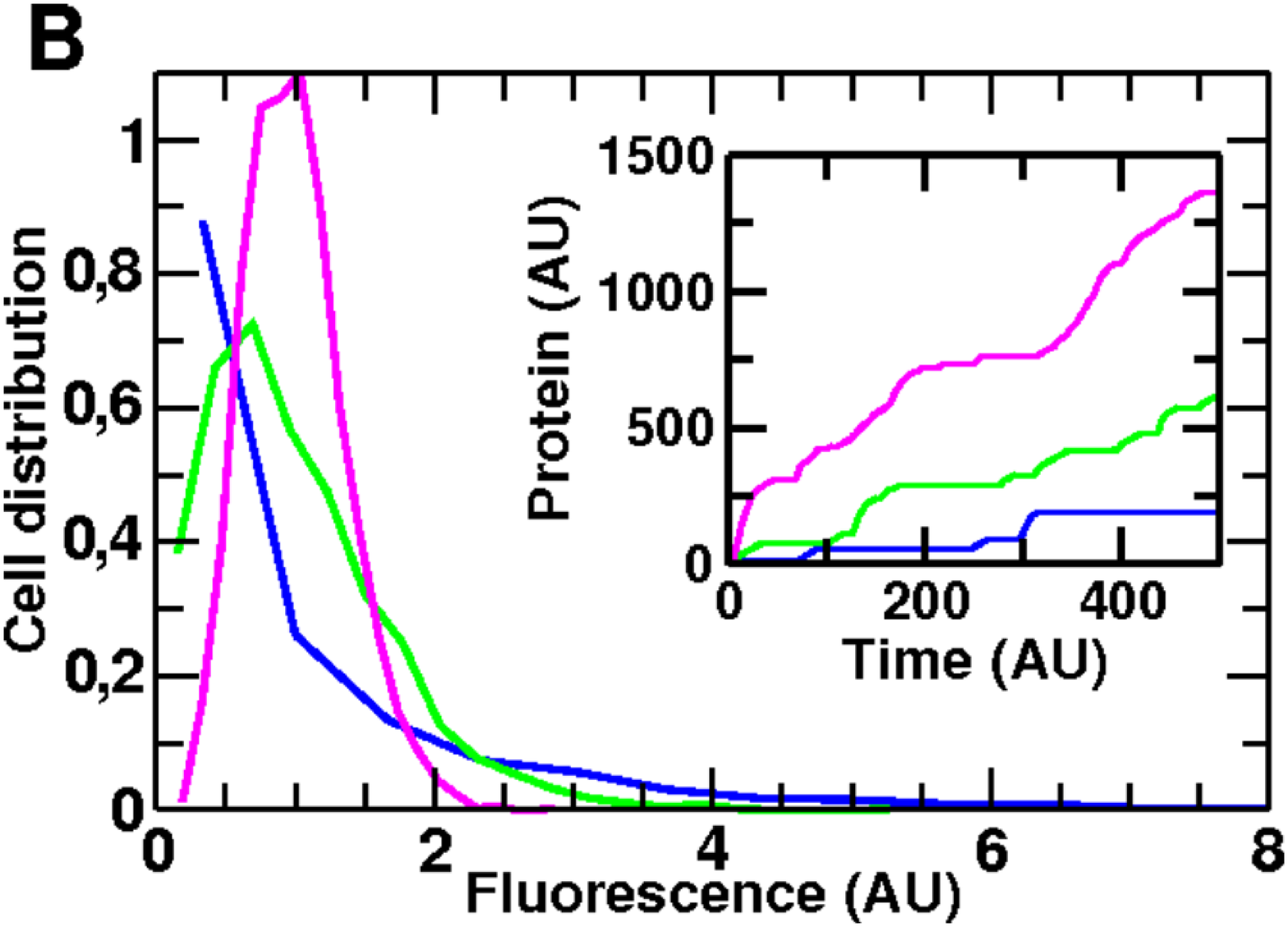}
\caption{\label{fig3}\textbf{A)} Simulations of different effects of activator binding on promoter switch-on/off rates for fixed and estimated parameters. Red and black: wild-type and \textit{flgM} cells of assuming that promoter activation and activator binding are independent (first scenario); green and blue: wild-type and \textit{flgM} cells assuming that the activator binding enhances promoter switch-on rate (second scenario); cyan and magenta dashed lines: wild-type and \textit{flgM} cells assuming activator binding decreases promoter switch-off rate (third scenario). Means, standard deviations and estimated parameters are summarized in Table~\ref{table1}, fixed parameters are found in the Materials and Methods. 
The inset shows the experimental levels of CheY, expressed as YFP fusion from native chromosomal position for wild-type (red) and \textit{flgM} cells (black). Redrawn from \cite{Kollmann05}. \textbf{B)} Different population distributions  (\textit{case I}), each normalized to mean 1 for varying promoter switch-on rates $\lambda_A^+$ at fixed switch-off rate $\lambda_A^-=0.1$. Mean activator binding: $ B_{eq}=\frac{1}{7}$. Blue: $\lambda_A^+=0.01$; green: $\lambda_A^+=0.05$; magenta: $\lambda_A^+=1$. The inset shows corresponding simulated protein trajectories for each switch-on rate $\lambda_A^+$.}
\end{center}
\end{figure}

In eukaryotes, promoter activation is believed to occur due to chromatin remodelling \cite{RaserOShea04} which erratically uncovers transcription-factor binding sites. Activator binding, however, is assumed to be quite fast and frequent, because of the high copy number of TFs. Therefore, independence of activator binding and promoter activation seems to be a reasonable assumption in eukaryotes. For prokaryotes, the situation is less clear since a possible explanation or mechanism for the slow process of promoter activation is still lacking, although it has been measured quite accurately \cite{Golding05,Cai06}. 
In recent experiments Elf et al. measured the time scale for the binding/unbinding of an activator/TF at the single-molecule level in a living cell of \textit{E. coli}. The binding/unbinding of highly abundant TFs is also suggested to be quite fast \cite{Elf07}. Therefore, activator binding does not seem to be the limiting step within the process of gene expression.
In order to gain insight into the influence of the activator binding on promoter activation in prokaryotes, we compare experimental data with simulations.
Kollmann et al. (see \cite{Kollmann05}, Fig.~2a; redrawn in Fig.~\ref{fig3}A, inset), compared the mean expression of CheY in a wild-type of \textit{E. coli} and \textit{flgM} cells, where the upstream transcription inhibitor, FlgM, was deleted. 
The deletion corresponds to a sevenfold over expression of CheY.
Several effects of an activator/repressor on the activation of the promoter are possible. We discuss the three most intuitive scenarios:\\
\textit{1. Activator binding and promoter activation are independent}: The RNA-polymerase can start transcription if and only if the activator is bound (repressor is unbound) \textit{and} the promoter is active.\\
\textit{2. Activator binding enhances promoter switch-on rate}: For an experimentally observed sevenfold over expression we assume that the switch-on rate $\lambda_A^+$ of the over expression of CheY is enhanced compared to that of the wild-type. The switch-off rate $\lambda_A^-$ is not affected.\\
\textit{3. Activator binding decreases promoter switch-off rate}: For an experimentally observed sevenfold over expression we assume that the switch-off rate $\lambda_A^-$ of the over expression is decreased compared to that of the wild-type. The switch-on rate $\lambda_A^+$ is not affected.\\
Of course, combinations of the mentioned scenarios are possible and likely to occur in nature. 
However, to keep the estimated parameters identifiable, we only focus on these three limiting scenarios.\\
The experimental data shows that the mean protein level of the over expressing \textit{flgM} cells is sevenfold higher than that of the wild-type cells. Furthermore, the standard deviation of the population distribution for the \textit{flgM} cells increases quite significantly compared to the wild-type cells (Fig.~\ref{fig3}A (inset) and Table~\ref{table1}).
Fig.~\ref{fig3}A shows the population distributions of the wild-type and \textit{flgM} cells for the different scenarios after parameter estimation (see Material and Methods). The parameters are estimated such that the wild-type standard deviation and the mean fluorescence level of the \textit{flgM} cells are represented best. The estimated parameters ($B_{eq}$ for the first scenario, $\lambda_A^+$, $\lambda_A^-$ and the level of the over expression $OE$ for second and third scenario) and the corresponding characteristics of the population distributions are summarized in Table~\ref{table1}.\\
The simulations reveal that there exits a set of parameters for the first scenario, where the activator binding does not influence the promoter activation process directly, such that the characteristic standard deviation of the wild-type and the mean fluorescence level of the \textit{flgM} cells is reproduced very well ($\chi^2=0.004$). Furthermore, it also mimics (without any optimization) the increased standard deviation of the \textit{flgM} cells (cf. Table~\ref{table1} and Fig. \ref{fig3}A, red line denotes wild-type, black line denotes sevenfold over expression). The mean activator binding for the \textit{flgM} cells is $7.3$ times larger than that of the wild-type cells ($B_{eq,WT}=0.13$, $B_{eq,OE}=0.95$) which leads to an about sevenfold protein over expression of the mean fluorescence level.
For the second scenario, where activator binding and promoter activation are not independent of each other, but activator binding enhances the promoter switch-on rate, the simulations with the estimated parameters does not represent the characteristic standard deviation of the wild-type and the mean fluorescence level of the \textit{flgM} cells equivalently well ($\chi^2=0.67$, Fig.~\ref{fig3}A green and blue lines denote wild-type and \textit{flgM} cells, resp.). The standard deviation of the \textit{flgM} cells becomes much larger than the experimental one.
The estimated parameters are given by $\lambda_A^+=0.005$ for the wild-type switch-on rate and $OE=413$ for the increased multiplication factor for the \textit{flgM} cells. The promoter is switched-on in 5$\%$ of the time for wild-type cells and 95$\%$ of the time for \textit{flgM} cells.
For the third scenario, where the binding of the activator decreases the promoter switch-off rate, a set of parameters can be found such that both characteristics are reproduced ($\chi^2=0.04$, Fig.~\ref{fig3}A cyan and magenta dashed lines represent wild-type and \textit{flgM} cells, resp.). The standard deviation of the \textit{flgM} cells is also increased quite well. The resulting parameters $\lambda_A^-=0.39$ for the wild-type switch-off rate and $OE=924$ for the reduction factor for the \textit{flgM} cells imply that the promoter is switched-off 90$\%$ of the time for wild-type and 1$\%$ of the time for \textit{flgM} cells.
If we compare the resulting skewness of the \textit{flgM} cells in each scenario with the experimentally measured one we find that the second scenario has the most positive skewness, but this scenario does not fit the characteristics quite well. 
For the first scenario, the characteristics are represented very well and the skewness is also increased compared to the third scenario.
We conclude that activator binding and promoter activation can be considered to good approximation as independent processes in prokaryotes. Of course, this hypothesis has to be investigated in further experiments.\\

Recently, stochastic dynamics has been linked to population distributions implying the classical model of burst-like transcription and translation \cite{Friedman06}. A Gamma-distribution fitted the stochastic simulations well and reproduced specific shapes of the population distribution at steady-state.\\
Here, we investigate the influence of the ratio $\kappa=\lambda_A^+/\lambda_A^-$ of the promoter switching rates on the shape of the population distribution.
Simulations of the model presented in Fig.~\ref{fig1} reveal that different cell distribution shapes can be generated for a fixed switch-off $\lambda_A^-$ rate by varying the switch on rate $\lambda_A^+$.
If the promoter activation is much smaller than its inactivation, $\kappa\ll 1$, the resulting protein distribution peaks at zero (Fig.~\ref{fig3}B, blue line).
For appropriate promoter activation rate, the protein distribution seems to be log-normally distributed (green line).
If the promoter activation rate exceeds the inactivation rate, $\kappa\geq 1$, the protein distribution will be shifted to a normal distribution (magenta).
The inset of Fig.~\ref{fig3}B shows the corresponding protein trajectories of the different switch-on rates.
The remaining reaction rates of activator binding, transcription and translation have an impact on the mean amount of protein or the noise strength, but not on the shape of the protein distribution itself.
If, however, the time scale of promoter activation is much faster than that of activator binding, e.g., for strongly repressed genes, the shape of the distribution can also be influenced by strong fluctuations in the activator concentration, i.e., the roles of the activator binding and promoter activation are interchanged.
Thus, the results of Fig.~\ref{fig3}B can also be obtained by fixing the promoter to its \textit{on} state and choosing appropriate parameters for the activator binding.

\section*{Discussion}
Different experiments have identified different causes for the main contribution to gene expression noise.
This implies that there might be no general rule for the main source of noise or comprehensive knowledge of the overall noise architecture. 
However, the theoretical framework presented in this study allows to quantify the relative contributions of the different sources of gene expression noise.
Taking cell division explicitly into account we derive time-dependent, non-equilibrium solutions for the mean amount of mRNA and protein as well as for the noise contributions from promoter activation and transcription.
In order to interpret the analytical results with respect to their biological relevance, we discuss asymptotic cases for the molecule lifetimes representing essential physiological cases.
Our analysis confirms the intuition that molecule lifetimes, compared to generation time, determine the influence of the noise contributions from previous generations on the overall noise level of the actual generation. Long-term memory effects and noise accumulation from previous generations might become important if molecule lifetimes are much larger than the generation time. 
In addition, the main contribution to the cell-to-cell variation within a clonal population depends strongly on the kinetic rates associated with the expression of each individual gene.
We show that the induction level of the activator or TF binding, $B_{eq}$, determines crucially the ratio of noise from promoter activation to noise from transcription, $\sigma_A^2/\sigma_R^2$, and thus the dominant source of noise.
It follows that the experimental set-up for the same organism \textit{S. cerevisiae} of \textit{in-vivo} experiments performed by Bar-Even et al. \cite{BarEven06} and \textit{in-vitro} experiments by Raser and O'Shea \cite{RaserOShea04}, plays a fundamental role for the experimentally measured noise level.
Low induced genes \cite{BarEven06} bear clear transcriptional noise signature due to low-copy number of mRNA molecules, whereas highly induced genes \cite{RaserOShea04} show typical characteristics of noise stemming from promoter activation (Fig.~\ref{fig2}). Therefore, the proposed model provides an explanation for these contradictory experimental results since it is able to reproduce both results for the same organism \textit{S. cerevisiae}. Additionally, this suggests that the incorporation of an activator binding process acting independent of the promoter activation process is important for a theoretical description of gene expression.\\
In eukaryotes, independence of activator binding and promoter activation is a reasonable assumption whereas in prokaryotes the situation is less clear. Based on published, experimentally measured cell distribution of wild-type and \textit{flgM} cells of \textit{E. coli}, we performed parameter estimation with the proposed model to discriminate between limiting cases of the effect of activator binding on promoter activation. Simulations reveal that for the first and third scenario a set of parameters can be found such that the characteristic standard deviation of the wild-type cells and mean fluorescence level of the \textit{flgM} cells can be reproduced quite well. An increased standard deviation for the \textit{flgM} cells can also be observed in both scenarios such that a qualitative distinction between both scenarios based on simulations seems to be difficult. 
The biological interpretation of the estimated parameters, however, suggest that the first scenario, i.e., the independence of promoter activation and activator binding, is more likely since the cellular effort of a sevenfold increase of the mean activator binding $B_{eq}$ is reasonable. 
If the binding of the activator increases/reduces the promoter switch-on/-off rate, in the second and third scenario, respectively, the promoter is switched-off most of the time (95$\%$/90$\%$) for the wild-type. 
However, these delay times for the protein production, even for a repressed gene, contradict the experimental observations in \textit{E. coli} where proteins are produced quite continuously (pers. communication V. Sourjik). 
Since the protein level should be strongly optimized to allow for reliable information processing, it seems to be very unlikely that promoters have evolved which are so strongly repressed that they are switched-off nearly all of the time.
Furthermore, the over expression factor $OE$ is estimated to be very large in the second ($OE=413$) and third ($OE=924$) scenario. This implies that an experimental depletion of an upstream inhibitor leading to a sevenfold increased mean fluorescence level corresponds to a theoretical $\approx$400/900fold increase/decrease of the switch-on/-off rate. The cellular effort to achieve this is expected to be very high, such that these scenarios seem to be inefficient.\\
The increased skewness (skewness of the \textit{flgM} cells of about 1.74) observed by Kollmann et al. in experiments is not reflected in any scenario of our underlying model. 
The proposed model is a reduced description of the overall system and does not take into account the complex flagella network such that an entire coincidence of the experiments and simulations is not expected.
The skewness might also be influenced by external factors, like variations in ribosome or polymerase concentrations or by
feedbacks of downstream to upstream processes, but none of these features are explicitly included in the present model.\\
Thus, the first scenario reproduces the experimental observations quite well and the estimated parameters can also be interpreted biologically reasonable. This observation implies that the promoter activation rates might be an intrinsic property of the biological system, both for the wild-type and the over expression line. One way to regulate gene expression is to finetune and control the mean binding of the activator.\\
If the time scales of activator binding and promoter activation can be separated, the ratio between the promoter activation and inactivation rate determines the shape of the population distribution (Fig.~\ref{fig3}B). This suggests that the often interpreted log-normal distribution of single cells does not result from consecutively multiplicative stochastic processes but rather reflects the switching rates of the slowest process within gene expression.\\

\section*{Materials and Methods}
\small
\textbf{Simulating stochastic processes.} We assume that activator binding and promoter activation can be described by a random telegraph process with transition rates $\lambda_B^+$, $\lambda_B^-$ (activator binding), $\lambda_A^+$ and $\lambda_A^-$ (promoter activation).
The initial state of the promoter is determined by drawing a uniformly distributed random number (URN) $r\in[0,1]$ and checking whether $r<\lambda_A^+/\gamma_A$, such that the promoter is \textit{on}. Otherwise it is \textit{off} in its initial state.
Transcription is a \textit{birth-and-death} process with time-dependent synthesis rate $\lambda_R^+(t)=\lambda_R^+A(t)B(t)$, i.e. mRNA can only be synthesized if the activator is bound \textit{and} the promoter is in its \textit{on}-state (see Results and Fig.~\ref{fig1}). 
The original Gillespie-algorithm \cite{Gillespie77} has been refined (for review see \cite{Gillespie06}), but also modified and extended to model growing cell volume via time-dependent reaction rates \cite{Lu04}. 
To determine the next time $\tau$ of the reaction and the next reaction $\mu$ itself for time-dependent reaction rates, we follow along the lines of Gillespie \cite{Gillespie77} and Lu et al. \cite{Lu04} and arrive at the cumulative distribution function 
\begin{equation}\label{Ftau}
F(\tau)=1-\exp\left[-\sum_{\mu}\int_0^{\tau}a_{\mu}(t+\tau')d\tau'\right]=:1-P_0(\tau).
\end{equation}
Drawing a URN $\bar{u}_1\in[0,1]$ we set $\bar{u}_1=1-P_0(\tau)$ and obtain, since $1-\bar{u}_1$ is also a URN, the new URN $u_1=P_0(\tau)$.
The stochastic time $\tau$ for the next reaction to occur is obtained by inverting this equation.
We formulate the cumulative distribution function for transcription with a time-dependent synthesis rate $a_1(t)=\lambda_R^+(t)=\lambda_R^+A(t)B(t)$ of mRNA. The degradation does not depend on time $\tau>t$ with transition rate $a_2=\lambda_R^-R(t)$ where $R(t)$ represents the actual amount of mRNA at time $t$.
We find that at time $t$, the next stochastic time $\tau$ has to satisfy
\begin{equation}\label{findtau}
\ln(u_1)=-\lambda_R^+\int_0^{\tau} B(t+\tau') A(t+\tau')d\tau'-\lambda_R^-R(t)\tau.
\end{equation}
Drawing a second URN $u_2\in[0,1]$, the next reaction $\mu$ must fulfill the inequality
\begin{equation}\label{nextmu}
\lambda_R^+A(t+\tau)B(t+\tau) < u_2\left(\lambda_R^+A(t+\tau)B(t+\tau)+\lambda_R^-R(t+\tau)\right)
\end{equation} 

\noindent The modified Gillespie-algorithm with time-dependent reaction rate $a_1(t)$ determines the next stochastic time $\tau$ in Eqn.~(\ref{findtau}) as the upper bound of the integral.
In our case, however, the integrand has a very special form, i.e., it is 1 if and only if the activator is bound as well as the promoter is \textit{on}. Otherwise the integrand is 0. 
Therefore, the integration becomes a simple summation over all \textit{on}-states, $O_{\tau}$, of the product of the activator times the promoter within the time interval $[t,t+\tau]$. We define $\alpha_{\tau}$ as the ratio of the \textit{on}-states to the time interval $[t,t+\tau]$, i.e., $0\leq\alpha_{\tau}=O_{\tau}/\tau\leq 1$. 
The next time $\tau$ can therefore be calculated according to Eqn.~(\ref{findtau}) which reduces for an exponentially distributed stochastic variable $\ln(u_1)=:-z$ to
\begin{equation}\label{findtauNew}
\tau=\frac{z}{\lambda_R^+\alpha_{\tau}+\lambda_R^-R(t)}.
\end{equation}
Note that $\alpha_{\tau}$ depends on the single realization of $A(t)B(t)$ and is thus also a stochastic variable.
If $A(t)$ and $B(t)$ are time-independent, e.g., $A(t)B(t)\equiv 1$, it follows that $\alpha_{\tau}=1$ and the original Gillespie-algorithm \cite{Gillespie77} is recovered.
Therefore, we use the original Gillespie-algorithm to determine the next time $\tau$ and reaction $\mu$ and check afterwards whether the proposed Gillespie-step can be performed or not. If $A(t+\tau)B(t+\tau)=1$, mRNA synthesis can be realized, but if $A(t+\tau)B(t+\tau)=0$ and mRNA synthesis is selected as reaction $\mu$, the step is rejected and new URNs are drawn. In general, this procedure will always select a stochastic time $\tau$ which is smaller than that of the modified algorithm of Eqn.~(\ref{findtauNew}) since $0\leq\alpha_{\tau}\leq 1$.
However, the above procedure of taking the original Gillespie-algorithm and rejecting specific reactions is equivalent to the determination of the next time $\tau$ via Eqn.~(\ref{findtauNew}). 
To obtain the same stochastic time for both procedures, the following equation should hold:
\begin{equation}
z=\frac{\lambda_R^+\alpha_{\tau}+\lambda_R^-R(t)}{\lambda_R^++\lambda_R^-R(t)}\bar z;
\end{equation}
where $z$ and $\bar z$ are exponentially distributed variable stemming from a URN $u$ via $\ln(u)$.
This equation is true since an exponentially distributed variable can be described by the product of a constant (given a specific $\alpha_{\tau}$) times another exponentially distributed stochastic variable $\bar z$.
Therefore, taking the original Gillespie-algorithm and rejecting specific reactions according to the time-dependent trajectory $A(t)B(t)$ is just another realization of the modified Gillespie from Eqn.~(\ref{findtauNew}) and averaging over a lot of trajectories yields the same result.\\
The generated mRNA trajectory $R(t)$ can directly be used to calculate the appropriate mean protein number $\langle X(t)\rangle_X$ given by Eqn.~(\ref{mean_n}).
At the end of one cell cycle, the cell divides symmetrically into two daughter cells.
The mother's amount of protein and mRNA is divided binomially to both daughter cells.
To demonstrate that the modified Gillespie-algorithm computes the correct solution, Fig.~S1 shows a comparison of simulated mRNA and protein trajectories with the analytical ones from Eqns.~(\ref{averagedmeanx}) and (\ref{averagedmeanr}), respectively.

\noindent\textbf{Generating population distributions.} In order to avoid dependence on the initial conditions, we start with a certain amount of mRNA and proteins and simulate in total 15 generations. 
After five generations we randomly determine one cell with its amount of mRNA and protein to be the mother cell for the next five generations. 
The system is equilibrated and the actual simulations can be started. 
We randomly choose one cell to be the mother cell and generate 10 offspring generations (=1024 cells in the 10th generation).
To compare the influence of the length of the generation time $T_G$ on the population distribution, we simulated populations with fixed generation time and varying generation time (choosing $T_G\in\mathcal{N}(\mu,\sigma)$).
We did not observe any significant differences between these two scenarios and therefore fix for simplicity the generation time in all simulations.

\noindent\textbf{Effect of activator binding on promoter activation: parameter estimation.} A least-square fit is performed with MATLAB such that, after data normalization to mean wild-type fluorescence of 1, the experimental standard deviation of the wild-type cells ($\sigma=0.69$) and mean \textit{flgM} fluorescence ($\mu=6.96$) of the population distributions are best represented.
For each optimization step, $20\times2^9=10240$ realizations of the proposed model are generated.
The following parameters are fixed for the simulations: $\lambda_R^+=2$, $\lambda_R^-=0.2$, $\lambda_X^+=4$ and $\lambda_X^-=10^{-4}$. In the first scenario, the promoter switch-on/-off rates are set to $\lambda_A^+=0.05$ and $\lambda_A^-=0.1$, representing realistic kinetic rates for a repressed gene \cite{Golding05}. The mean activator binding rates of wild-type ($B_{eq,WT}$) and \textit{flgM} cells ($B_{eq,OE}$) are estimated separately.
For the second and third scenario we assume a mean activator binding of $B_{eq}=0.5$ for the wild-type as well as for the \textit{flgM} cells.
The promotor switch-off rate is set to $\lambda_A^-=0.1$ and the promoter switch-on $\lambda_{A,WT}^+$ for the wild-type and the strength of the over expression, $OE$, are estimated in the second scenario. If the activator is bound, the promoter switch-on rate for the \textit{flgM} cells, $\lambda_{A,OE}^+$, will be enhanced by the over expression factor $OE$, i.e. $\lambda_{A,OE}^+=OE\times\lambda_{A,WT}^+$.  In the third scenario we set the promoter switch-on rate to $\lambda_A^+=0.05$ and estimate the promoter switch-off rate $\lambda_{A,WT}^-$ for the wild-type and the strength of the over expression $OE$. If the activator is bound, the promoter switch-off rate for the \textit{flgM} cells, $\lambda_{A,OE}^-$, will be reduced by the over expression factor $OE$, i.e. $\lambda_{A,OE}^-=\lambda_{A,WT}^-/OE$. 
The estimated parameters are summarized in Table~\ref{table1}.

\normalsize
\section*{Acknowledgements}
This work was financially supported by the Graduate School ``Plant Signal Systems'' (DFG grant no. GK1305 for JR) and the DFG Emmy Noether-Program (for MK).\\
\textbf{Competing interests} The authors have declared that no competing interests exist.\\
\textbf{Author's contributions} MK initiated and supervised the study, MK and JR carried out the analytical calculations, JR performed the simulations, JR, CF and MK wrote the manuscript with useful comments from JT.\\
\textbf{Abbreviations} G/C/YFP, green/cyan/yellow fluorescent protein; TF, transcription factor; OE, over expression;  URN, uniform random number

\newpage

\begin{table}
\begin{tabular}{c|c|c|c|c|c|c}
 \multicolumn{2}{c|}{} & & & & estimated & \\ 
 \multicolumn{2}{c|}{} & $\mu$ & $\sigma$ & skewness & parameter & $\chi^2$ \\ \hline\hline
Experiment & wild-type & 1 & 0.69 & 2.51 & & \\ \cline{2-5}
& \textit{flgM} cells & 6.96 & 3.38 & 1.74 & & \\ \hline\hline
Simulation & wild-type & 1 & 0.67 & 1.26 & $B_{eq}=0.13$ & \\ \cline{2-6}
1st scenario & \textit{flgM} cells & 7.02 & 2.63 & 0.44 & $B_{eq}=0.94$ & 0.004\\ \hline
Simulation & wild-type & 1 & 1.51 & 2.86 &  $\lambda_A^+=0.005$ & \\ \cline{2-6}
2nd scenario & \textit{flgM} cells & 6.92 & 5.25 & 1.07 &  $OE=413$ & 0.67 \\ \hline
Simulation & wild-type & 1 & 0.63 & 1.17 & $\lambda_A^-=0.39$ & \\ \cline{2-6}
3rd scenario & \textit{flgM} cells & 6.77 & 2.56 & 0.1 & $OE=924 $& 0.04\\ \hline
\end{tabular}
\caption{\label{table1} Characteristic mean ($\mu$), standard deviation ($\sigma$) and skewness for population distributions from experiments of Kollmann et al. \cite{Kollmann05} and simulations using the proposed stochastic model. Cells in which the upstream transcription inhibitor, the anti-sigma factor FlgM, was deleted, are denoted by \textit{flgM} cells. The deletion leads to a sevenfold over expression. Differences in the simulated scenarios, interpretation of the factor $OE$ and parameter estimation are described in the text and in Material and Methods.}
\end{table}

\end{document}


\renewcommand{\thetable}{S\arabic{table}}
\renewcommand{\theequation}{S\arabic{equation}}
\renewcommand{\thefigure}{S\arabic{figure}}

\begin{center}
\textbf{\Large Supporting Information}
\end{center}

\section{Derivation of time-dependent solution for transcription}\label{sec1}

For an activated promoter, the process of transcription is assumed to be Poissonian.\\ 
Assume that $R(t)$ is the number of mRNA transcribed from an activated promoter $A(t)B(t)=1$.
The stochastic variables $A(t),B(t) \in \{0,1\}$ denote the promoter activation and activator binding, respectively. 
We assume $R(t)\stackrel{\lambda_R^+}{\longrightarrow}R(t)+1$ with time-dependent transition rate $\lambda_R^+(t)=\lambda_R^+A(t)B(t)$ and for the inverse process $R(t)\stackrel{\lambda_R^-}{\longrightarrow}R(t)-1$ where $\lambda_R^-$ is the degradation rate.\\

Assuming initially $R_0$ molecules at time $t_0$, the stochastic Master equation for this Poisson process has the form 
\begin{eqnarray}
\partial_tP(R,t|R_0,t_0)&=&\lambda_R^+(t)P(R-1,t|R_0,t_0)+(R+1)\lambda_R^-P(R+1,t|R_0,t_0)\nonumber\\
&&\qquad\qquad\qquad\qquad\qquad-(R\lambda_R^-+\lambda_R^+(t))P(R,t|R_0,t_0).\label{mastertransc}
\end{eqnarray}

To solve the equation we define the generating function
\begin{eqnarray}
G(s,t|R_0,t_0)&=&\sum_R s^RP(R,t|R_0,t_0)	\mbox{  such that} \nonumber \\
\partial_t G(s,t|R_0,t_0)&=& \lambda_R^+(t)(s-1)G(s,t|R_0,t_0)-\lambda_R^-(s-1)\partial_sG(s,t|R_0,t_0). \label{partial_tG}
\end{eqnarray} 
We choose the ansatz
\begin{equation}
G(s,t)=e^{(s-1)\alpha(t)}\psi(s,t)
\end{equation}
for which we obtain using (\ref{partial_tG})
\begin{eqnarray}
\partial_t G(s,t)&=&\partial_t\alpha(t)(s-1)e^{(s-1)\alpha(t)}\psi(s,t)+e^{(s-1)\alpha(t)}\partial_t\psi(s,t) \nonumber\\
&=& \lambda_R^+(t)(s-1)e^{(s-1)\alpha(t)}\psi(s,t)-\lambda_R^-(s-1)e^{(s-1)\alpha(t)}[\alpha(t)\psi(s,t)+\partial_s\psi(s,t)] \nonumber\\
&=&(\lambda_R^+(t)-\lambda_R^-\alpha(t))(s-1)e^{(s-1)\alpha(t)}\psi(s,t) -\lambda_R^-(s-1)\partial_s\psi(s,t)e^{(s-1)\alpha(t)}.
\end{eqnarray}
Comparison of the coefficients yields
\begin{eqnarray}
\partial_t\alpha(t)&=&\lambda_R^+(t)-\lambda_R^-\alpha(t) \label{alpha}\\
\partial_t\psi(s,t) &=& -\lambda_R^-(s-1)\partial_s\psi(s,t). \label{partial_tpsi}
\end{eqnarray}
The solution of (\ref{alpha}) is given by
\begin{equation}\label{alphasolution}
\alpha(t)=\alpha(t_0)e^{-\lambda_R^-(t-t_0)}+\int_{t_0}^t e^{-\lambda_R^-(t-t')}\lambda_R^+(t')dt' .
\end{equation}

To determine $\psi(s,t)$, we introduce a new variable $z:=\ln(s-1)$ (with $dz=\frac{1}{s-1}ds$), substitute $s$ by $z$ in (\ref{partial_tpsi}) and obtain the differential equation
\begin{equation}
\partial_t\tilde{\psi}(z,t)=-\lambda_R^-\partial_z\tilde{\psi}(z,t).
\end{equation}
Its solution is an arbitrary function $F$ of the variable $\lambda_R^-t-z$, so we can write $\tilde{\psi}(z,t)=F[e^{z-\lambda_R^-t}]$
and therefore (re-substituting $z$ by $s$)
\begin{equation}
\psi(s,t)= F[(s-1)e^{-\lambda_R^-t}].
\end{equation}
It follows for the generating function 
\begin{equation} \label{genfunct}
G(s,t)=F[(s-1)e^{-\lambda_R^-t}]e^{(s-1)\alpha(t)}.
\end{equation}
Normalization requires $G(1,t|R_0,t_0)=1$ and therefore $F[0]=1.$\\
For a complete determination of $F$, we have to include the initial condition $P(R,0|R_0,0)=\delta_{RR_0}$ which implies
\begin{equation}
G(s,0|R_0,t_0)=\sum_Rs^R\underbrace{P(R,0|R_0,0)}_{=\delta_{RR_0}}=s^{R_0}=F[s-1]e^{(s-1)\alpha(0)}.
\end{equation}
We arrive at $F[s-1]=[(s-1)+1]^{R_0} e^{-(s-1)\alpha(0)}$ which gives for the substitution $x:=s-1$
\begin{equation}
F[x]=[x+1]^{R_0} e^{-x\alpha(0)}. \label{F}
\end{equation} 
Replacing $F$ in (\ref{genfunct}) with our fully determined $F$ of (\ref{F}) we obtain

\begin{equation} 
G(s,t|N,0) =[(s-1)e^{-\lambda_R^-t}+1]^N \exp[(s-1)(\alpha(t)-\alpha(0)e^{-\lambda_R^-t})] \label{completegenfunct}
\end{equation} 

To start at an arbitrary time $t_0$ and molecule number $R_0$ we now include the effect of cell division into the generating function Eqn.~(\ref{completegenfunct}) of the transcription process.
It is obvious that the noise level strongly depends on the mean number of mRNAs.
If there were many mRNA copies the noise from cell division would become smaller, such that the process of cell division would not account for the variability among daughter cells.
Conversely, only a few copies of mRNA result in a large variability among daughter cells.
Therefore this effect has to be taken into account when focussing on gene expression noise.\\
Activator binding and promoter activation are expected to be fairly independent on cell division and the amount of protein is simply halved if the cell divides symmetrically as for most systems the copy number of proteins is $>10^2$.
Thus transcription is the only process where we have to account for cell division in the probability density function.

Assuming exactly one symmetric cell division with binomial distribution of the mRNA at time $t_1$ within $t_0<t_2$, the expression for the conditional probability density function of mRNA synthesis and decay is given by
\begin{equation} \label{probbinomial}
P(R_2^-,t_2|R_0^+,t_0)=\sum_{R_1^-=0}^{\infty}\sum_{R_1^+=0}^{R_1^-} P(R_2^-,t_2|R_1^+,t_1){R_1^-\choose R_1^+}2^{-R_1^-} P(R_1^-,t_1|R_0^+,t_0),
\end{equation}
with $R_1^+$ and $R_1^-$ is the amount of mRNA of the daughter and mother cell, respectively.\\
It can be easily seen via induction over the number of cell divisions, that the probability density function of mRNA synthesis and decay after $n$ cell divisions at $t_1,\ldots,t_{n}$ reads with $t_{n+1}>t>t_n$ and $t_{1}>t'>t_0$ 
\begin{eqnarray} 
P(R,t|R',t')&=&\sum_{R_{n}^-=0}^{\infty}\sum_{R_{n}^+=0}^{R_{n}^-} P(R,t|R_{n}^+,t_{n}){R_{n}^-\choose R_{n}^+}2^{-R_{n}^-} P(R_{n}^-,t_{n}|R',t') \nonumber\\
&=&\sum_{R_{n}^-=0}^{\infty}\sum_{R_{n}^+=0}^{R_{n}^-}\cdots\sum_{R_{1}^-=0}^{\infty}\sum_{R_{1}^+=0}^{R_{1}^-} P(R,t|R_{n}^+,t_{n}){R_{1}^-\choose R_{1}^+}2^{-R_{1}^-}P(R_1^-,t_1|R',t')\nonumber\\
&&\times\prod_{i=2}^{n}{R_{i}^-\choose R_{i}^+}2^{-R_{i}^-}P(R_{i}^-,t_{i}|R_{i-1}^+,t_{i-1}). \label{propfunctncelldiv} 
\end{eqnarray}
The generating function corresponding to one cell division at $t_1$ is given by
\begin{eqnarray} 
G(s_2,t_2|R_0,t_0)&=&\sum_{R_2,R_1^+,R_1^-} s_2^{R_2}\underbrace{P(R_2,t_2|R_1^+,t_1){R_1^-\choose R_1^+}2^{-R_1^-} P(R_1^-,t_1|R_0,t_0)}_{=P(R_2,t_2|R_0,t_0)}\\
   &\stackrel{(\ref{completegenfunct})}{=}&\sum_{R_1^+,R_1^-}[(s_2-1)e^{-\lambda_R^-(t_2-t_1)}+1]^{R_1^+} \exp\left[(s_2-1)\int_{t_1}^{t_2}e^{-\lambda_R^-(t_2-t')}\lambda_R^+(t')dt'\right]\nonumber\\
   &&\times{R_1^-\choose R_1^+}2^{-R_1^-} P(R_1^-,t_1|R_0,t_0)\nonumber
\end{eqnarray}
\begin{eqnarray}
&=&\sum_{R_1^-}\sum_{R_1^+=0}^{R_1^-}[(s_2-1)e^{-\lambda_R^-(t_2-t_1)}+1]^{R_1^+}{R_1^-\choose R_1^+}2^{-R_1^-} \nonumber\\
&&\times\exp\left[(s_2-1)\int_{t_1}^{t_2}e^{-\lambda_R^-(t_2-t')}\lambda_R^+(t')dt'\right] P(R_1^-,t_1|R_0,t_0)\nonumber\\
&=&\sum_{R_1^-}\bigg[\underbrace{\frac{1}{2}(s_2-1)e^{-\lambda_R^-(t_2-t_1)}+1}_{s_1}\bigg]^{R_1^-}\nonumber\\ &&\times\exp\left[(s_2-1)\int_{t_1}^{t_2}e^{-\lambda_R^-(t_2-t')}\lambda_R^+(t')dt'\right]P(R_1^-,t_1|R_0,t_0)\\
&\stackrel{(\ref{completegenfunct})}{=}&\bigg[(s_1-1)e^{-\lambda_R^-(t_1-t_0)}+1\bigg]^{R_0}\exp\left[(s_2-1)\int_{t_1}^{t_2}e^{-\lambda_R^-(t_2-t')}\lambda_R^+(t')dt'\right]\nonumber\\
&&\times\exp\left[(s_1-1)\int_{t_0}^{t_1}e^{-\lambda_R^-(t_1-t')}\lambda_R^+(t')dt'\right]\\
&=&\bigg[\frac{1}{2}(s_2-1)e^{-\lambda_R^-(t_2-t_0)}+1\bigg]^{R_0}\exp\left[(s_2-1)\int_{t_1}^{t_2}e^{-\lambda_R^-(t_2-t')}\lambda_R^+(t')dt'\right]\nonumber\\
&&\times\exp\left[\frac{1}{2}(s_2-1)\int_{t_0}^{t_1}e^{-\lambda_R^-(t_2-t')}\lambda_R^+(t')dt'\right].\label{onecelldiv}
\end{eqnarray}
By repeated application of the method above, i.e. assuming a finite number of cell divisions, the generating function is given by
\begin{eqnarray}
G(s,t|R_0,t_0)&=&\left[2^{-D(t,t_0)}(s-1)e^{-\lambda_R^-(t-t_0)}+1\right]^{R_0} \nonumber\\
&&\times\exp\left[(s-1)\int_{t_0}^{t}\lambda_R^+(t')e^{-\lambda_R^-(t-t')}2^{-D(t,t')}dt'\right]\nonumber\\
&=:& (\tilde{x}+1)^{R_0}K(s,t) \label{genfuncelldiv}
\end{eqnarray}
where $D(t,t')=D(t',t)$ represents the number of cell divisions within the time interval $[t,t']$.
The above formula is easily proven via induction over the number $D(t,t')$ of cell divisions within $[t,t']$.

We first compute the mean mRNA copy number, $\langle R(t)\rangle_R$, with the help of the generating function Eqn. (\ref{genfuncelldiv}). 
Setting $t_0=-\infty$, i.e. including an infinite number of cell divisions, we find that
\begin{equation}
G(s,t|R_0,-\infty)=\exp\left[(s-1)\alpha(t)\right] \label{Ginfty}
\end{equation}
with $\alpha(t)$ defined by
\begin{eqnarray}
\alpha(t)&=&\int_{-\infty}^{t}\lambda_R^+(t')e^{-\lambda_R^-(t-t')}2^{-D(t,t')}dt'\nonumber\\
         &=&\int_{-\infty}^{t_0}\lambda_R^+(t')e^{-\lambda_R^-(t-t')}2^{-D(t,t')}dt'+\int_{t_0}^{t}\lambda_R^+(t')e^{-\lambda_R^-(t-t')}2^{-D(t,t')}dt'\nonumber\\
&=&e^{-\lambda_R^-(t-t_0)}2^{-D(t,t_0)}\alpha(t_0)+\int_{t_0}^{t}\lambda_R^+(t')e^{-\lambda_R^-(t-t')}2^{-D(t,t')}dt'.
\end{eqnarray}
Note that $D(t,t')=D(t,t'')+D(t'',t')$ and that $\alpha(t)$ introduced here is a generalization of $\alpha(t)$ without cell division (cf.(\ref{alphasolution})).\\

As Eqn. (\ref{genfuncelldiv}) is the generating function of a Poisson process we obtain the mean of $R(t)$ by calculating $\partial_s G|_{s=1}$
\begin{equation}
\alpha(t)=\langle R(t) \rangle_R =\langle R(t_0) \rangle_R 2^{-D(t_0,t)}e^{-\lambda_R^-(t-t_0)}+\int_{t_0}^t\lambda_R^+(t')e^{-\lambda_R^-(t-t')}2^{-D(t,t')}dt'.\label{meanR}\\
\end{equation}

To calculate the variance we use a generating function $\Theta(s,t)$ which contains the joint probability function $P(R,t;R_0,t_0)$ representing the probability that at time $t$ there are $R$ molecules and at time $t_0$ there are $R_0$ molecules.
Afterwards we determine the variances by
\begin{equation}
\langle R(t)R(t_0)\rangle_R=\partial_s\Theta|_{s=1}.
\end{equation}
We again assume that initially the process is Poissonian-distributed and use the notation of (\ref{genfuncelldiv}), i.e. $G(s,t|R_0,t_0)=(x+1)^{R_0}K(s,t)$.
The generating function $\Theta(s,t)$ is given by 
\begin{eqnarray}
\Theta(s,t)&:=&\sum_{R_0} R_0 \sum_Rs^RP(R,t;R_0,t_0)=\sum_{R_0} R_0 \underbrace{\sum_Rs^RP(R,t|R_0,t_0)}_{G(s,t|R_0,t_0)}P(R_0,t_0) \nonumber \\
&=&\sum_{R_0} R_0 (x+1)^{R_0} K(s,t)\frac{e^{-\alpha(t_0)}\alpha(t_0)^{R_0}}{R_0!} \nonumber\\
&=& \sum_{R_0}(x+1)K(s,t)\partial_x(x+1)^{R_0}\frac{e^{-\alpha(t_0)}\alpha(t_0)^{R_0}}{R_0!} \nonumber\\
&=&(x+1)K(s,t)\partial_xe^{-\alpha(t_0)}\underbrace{\sum_{R_0}\frac{(\alpha(t_0)(x+1))^{R_0}}{R_0!}}_{=e^{\alpha(t_0)(x+1)}} \nonumber\\
&=&(x+1)K(s,t)\alpha(t_0)e^{\alpha(t_0)x} \nonumber 
\end{eqnarray}
\begin{eqnarray}
&=&[2^{-D(t_0,t)}(s-1)e^{-\lambda_R^-(t-t_0)}+1]\alpha(t_0)\nonumber\\
 &&\times\exp\bigg[(s-1)\big(\alpha(t_0)2^{-D(t_0,t)}e^{-\lambda_R^-(t-t_0)}+\int_{t_0}^t\lambda_R^+(t')e^{-\lambda_R^-(t-t')}2^{-D(t,t')}dt'\big)\bigg] \nonumber \\
&=&[2^{-D(t_0,t)}(s-1)e^{-\lambda_R^-(t-t_0)}+1]\alpha(t_0)\exp[(s-1)\langle R(t)\rangle_R].
\end{eqnarray}

It follows that 
\begin{equation}
\langle R(t)R(t_0)\rangle_R=\partial_s\Theta|_{s=1} =\alpha(t_0)(2^{-D(t_0,t)}e^{-\lambda_R^-(t-t_0)}+\langle R(t)\rangle_R).
\end{equation}
Replacing $t_0$ by $t'$ we obtain for the time correlation
\begin{eqnarray} \label{timecorrR}
\langle R(t)R(t')\rangle_R &=& \alpha(t')(2^{-D(t',t)}e^{-\lambda_R^-(t-t')}+\langle R(t)\rangle_R) \nonumber\\
&=&\langle R(t')\rangle_R 2^{-D(t',t)}e^{-\lambda_R^-(t-t')}+\langle R(t)\rangle_R\langle R(t')\rangle_R. 
\end{eqnarray}

\section{Partial Contributions to Gene Expression Noise}\label{sec2}

For a one-gene system (cf. Fig.~1 main text) whose gene product does not interfere with external parameters (e.g. expression for genes forming ribosomes do infer with the translation rate and thus with external parameters) the total variation $\sigma_{tot}^2$ can be written as a sum of the following contributions 
\begin{eqnarray}
\sigma_{tot}^2=\sigma_I^2 + \sigma^2_E.
\end{eqnarray}
The term $\sigma_I^2$ is the intrinsic noise contribution for fixed environmental conditions whereas the term $\sigma_E^2$ represents the effect of extrinsic noise on the mean expression level. 
The above expression for intrinsic noise is defined by
\begin{eqnarray}
\sigma_I^2&=&\left\langle \left(X(t)-\langle X(t)\rangle_{X,R,A,B}\right)^2\right\rangle_{X,R,A,B}\\
&=& \left\langle \left(X(t)+\langle X(t)\rangle_X-\langle X(t)\rangle_X -\langle X(t)\rangle_{X,R,A,B}\right)^2\right\rangle_{X,R,A,B}\label{a1}\\
&\stackrel{\langle\rangle_X}{=}& \left\langle \left(X(t)-\langle X(t)\rangle_X\right)^2\right\rangle_{X,R,A,B} + \left\langle \left(\langle X(t)\rangle_X-\langle X(t)\rangle_{X,R,A,B}\right)^2\right\rangle_{R,A,B} \nonumber\\
&+&\underbrace{\langle 2(X(t)-\langle X(t)\rangle_X)(\langle X(t) \rangle_X-\langle X(t)\rangle_{X,R,A,B})\rangle_{X,R,A,B}}_{=0,\mbox{ after } \langle\, .\, \rangle_X} \label{a2}\\
&=&\left\langle \left(X(t)-\langle X(t)\rangle_X\right)^2\right\rangle_{X,R,A,B} + \left\langle \left(\langle X(t)\rangle_X-\langle X(t)\rangle_{X,R}\right)^2\right\rangle_{R,A,B} \nonumber\\
&+& \left\langle \left(\langle X(t)\rangle_{X,R}-\langle X(t)\rangle_{X,R,A}\right)^2\right\rangle_{A,B} + \left\langle \left(\langle X(t)\rangle_{X,R,A}-\langle X(t)\rangle_{X,R,A,B}\right)^2\right\rangle_{B}\\
&=&\sigma^2_X+\sigma^2_R+\sigma^2_A+\sigma^2_B\label{noisecont}
\end{eqnarray}
where the methods of Eqns.~(\ref{a1}) and (\ref{a2}) are applied iteratively to the remaining terms.
The joint probability $P(X(t);R(t);A(t);B(t))$ can be written in terms of the following conditional probabilities, $P(X(t);R(t);A(t);B(t))=P(X(t)|R(t))P(R(t)|A(t),B(t))P(A(t))P(B(t))$. The averaging over the processes is given by $\langle X(t)\rangle_X=\int X(t)P(X(t)|R(t))dX(t)$ where $\int P(X(t)|R(t))dX(t)=1$.\\
The resulting terms of Eqn.~(\ref{noisecont}) give the second moment contributions from translation, $\sigma_X^2$, transcription, $\sigma_R^2$, promoter activation, $\sigma_A^2$, and activator binding, $\sigma_B^2$. Contribution from extrinsic noise consist of extrinsic effects on the mean expression and an increase of intrinsic noise due to extrinsic factors.
The above contributions can be calculated analytically as shown in the following.\\ 

As mentioned in the main text, we assume a separation of the time scales of activator binding and promoter activation, such that we assume an equilibrated activator binding process. To make the calculations feasible, we additionally assume a fixed induction level of the activator, $B_{eq}$. 
We use the following abbreviations, where $T_G$ is the generation time, $\alpha$ denotes either $X$ or $R$, $\beta$ represents either $A$ or $E$ and $n\in\mathbb N$.

\begin{eqnarray}
Q_X&=&\frac{\lambda_X^+\lambda_R^+}{\lambda_X^-\lambda_R^-}\langle A \rangle\\\
{\cal A}&=&\langle A \rangle B_{eq}\lambda_R^+\lambda_X^+\\
Z_{\alpha}^{(n)}(T_G)&=&\frac{1-\exp[-n\lambda_{\alpha}^-T_G]}{1-2^{-n}\exp[-n\lambda_{\alpha}^-T_G]}\\
Z_{\beta}^{(n)}(T_G)&=&\frac{1-\exp[-n\gamma_{\beta}^-T_G]}{1-2^{-n}\exp[-n\gamma_{\beta}^-T_G]}\\
Z_{\alpha\beta}(T_G)&=&\frac{1-\exp[(\lambda_{\alpha}^-+\gamma_{\beta}^+)T_G]}{1-2^{-1}\exp[(\lambda_{\alpha}^-+\gamma_{\beta}^+)T_G]}.
\end{eqnarray}

The expression levels within a large population are calculated after infinitely many cell divisions. The cell division times are given by $t_i$; by $i=0$ we define the last cell division that has occurred. Furthermore, we denote by $t_i^-$ and $t_i^+$ the time just before and after the cell division, respectively. Given the $n$-th moment of the stochastic function for mRNA, $\mu_R^{(n)}(t)$, and protein number, $\mu_X^{(n)}(t)$, the following invariance conditions can be derived
\begin{eqnarray}
\mu_R^{(n)}(t)=\mu_R^{(n)}(t+kT_G)   \qquad \forall \quad  k \in \mathbb Z\\
\mu_X^{(n)}(t)=\mu_X^{(n)}(t+kT_G)   \qquad \forall \quad k \in \mathbb Z.
\end{eqnarray}

Before the next cell division at time $t_1$, the mean mRNA and mean protein number of a large population is given by 
\begin{eqnarray}
\langle X(t) \rangle_{X,R,A}&=&\langle \hat X(t-t_0^+)\rangle_{X,R,A} + \frac{1}{2}e^{-\lambda_X^-(t-t_0)}\langle X(t_0^-) \rangle_{X,R,A}\nonumber\\
&=&\langle \hat X(t-t_0^+)\rangle_{X,R,A}+ \frac{2^{-1}}{1-2^{-1}e^{-\lambda_X^-T_G}}e^{-\lambda_X^-(t-t_0)}\langle\hat X(T_G)\rangle_{X,R,A}\\
\langle R(t) \rangle_{R,A}&=&\langle \hat R(t-t_0^+)\rangle_{R,A} + \frac{1}{2}e^{-\lambda_R^-(t-t_0)}\langle R(t_0^-) \rangle_{R,A}\nonumber\\
&=&\langle \hat R(t-t_0^+)\rangle_{R,A}+ \frac{2^{-1}}{1-2^{-1}e^{-\lambda_R^-T_G}}e^{-\lambda_R^-(t-t_0)}\langle\hat R(T_G)\rangle_{R,A}.
\end{eqnarray}
Here $ \hat X(t-t_0^+)$ represents the amount of protein resulting from synthesis in an actual generation, whereas $X(t_0^-)$ is the amount of protein in the mother cell just before cell division. The precise expressions for $\hat X(t-t_0^+)$ and $\hat R(t-t_0^+)$ are given by
\begin{eqnarray}
\langle \hat X(t-t_0^+)\rangle_{X,R,A}&=&\frac{{\cal A}}{\lambda_R^-} \left[\frac{1}{\lambda_X^-}\left(1-e^{-\lambda_X^-(t-t_0^+)}\right) - \frac{1-2^{-1}Z_R^{(1)}}{\lambda_R^--\lambda_X^-}\left(e^{-\lambda_X^-(t-t_0^+)}-e^{-\lambda_R^-(t-t_0^+)} \right)  \right]\nonumber\\
\langle \hat R(t-t_0^+) \rangle_{R,A}&=& \langle A \rangle B_{eq}\frac{\lambda_R^+}{\lambda_R^-}  \left[1-e^{-\lambda_R^-(t-t_0^+)}\right].\nonumber
\end{eqnarray}
We obtain the expressions for the mean amount of protein and mRNA given in Eqns.~(13), (14) in the main text.

The \textbf{transcriptional contribution} is given by
\begin{eqnarray}
&&\left\langle \left(\langle X(t) \rangle_X - \langle X(t) \rangle_{X,R}\right)^2 \right\rangle_{R,A}\\ &=& 
\underbrace{\left\langle \left(\langle X(t_0^+) \rangle_X - \langle X(t_0^+) \rangle_{X,R}\right)^2 \right\rangle_{R,A}e^{-2\lambda_X^-(t-t_0)}}_{T_1}\\
&+& \underbrace{2\int_{t_0}^t \left\langle \langle X(t_0^+) \rangle_X \lambda_X(t') - \langle X(t_0^+) \rangle_{X,R}\langle \lambda_X(t') \rangle_R \right\rangle_{R,A}e^{-\lambda_X^-(t-t_0)}e^{-\lambda_X^-(t-t')}dt'}_{T_2} \label{T2trans}\\
&+& \underbrace{\int_{t_0}^t\int_{t_0}^t \left\langle \lambda_X(t') \lambda_X(t'') - \langle \lambda_X(t') \rangle_R\langle \lambda_X(t'') \rangle_R \right\rangle_{R,A}e^{-\lambda_X^-(t-t'')}e^{-\lambda_X^-(t-t')}dt'dt''}_{T_3} \label{T3trans}.
\end{eqnarray}

From the invariance $t\rightarrow t+T_G$ we get an expression for the first term
\begin{eqnarray}
T_1(t-t_0)&=&\frac{1}{4}\frac{T_2(T_G)+T_3(T_G)}{1-\frac{1}{4}e^{-2\lambda_X^-T_G}}e^{-2\lambda_X^-(t-t_0)}.
\end{eqnarray}
Correlations between actual and previous generations are defined by
\begin{eqnarray}
T_2(t-t_0)&=& \langle A \rangle\frac{\lambda_R^+}{\lambda_R^-} (\lambda_X^+)^2 B_{eq} \frac{1}{2} \left[1-\frac{1}{4}e^{-(\lambda_R^-+\lambda_X^-)T_G}\right]^{-1}\left\{\frac{1}{\lambda_R^-+\lambda_X^-}\left[1- e^{-(\lambda_R^-+\lambda_X^-)T_G} \right]\right. \label{T2transana} \\
&-& \left.\frac{1}{\lambda_X^-}\left[1-e^{-\lambda_X^-T_G}  \right]e^{-\lambda_R^-T_G}(1-2^{-1}Z_R^{(1)})\right\}\frac{1}{\lambda_R^--\lambda_X^-}\left[e^{-2\lambda_X^-(t-t_0)}-e^{-(\lambda_R^-+\lambda_X^-)(t-t_0)}\right].\nonumber
\end{eqnarray}
Noise from the new generation is given by
\begin{eqnarray}
T_3(t-t_0)&=&2 Q_X B_{eq} \left\{(1-2^{-1}Z_R^{(1)})\left[-\frac{\lambda_X^+}{\lambda_R^--2\lambda_X^-}\left(e^{-2\lambda_X^-(t-t_0)}-\ e^{-\lambda_R^-(t-t_0)}\right) \right.\right. \label{T3transana} \\
 &+&\left.\left. \frac{\lambda_X^+}{\lambda_R^--\lambda_X^-}\left(e^{-2\lambda_X^-(t-t_0)}- e^{-(\lambda_X^-+\lambda_R^-)(t-t_0)}\right)\right] + \frac{\lambda_X^+}{2(\lambda_R^-+\lambda_X^-)}\left(1-e^{-2\lambda_X^-(t-t_0)}\right)\right.\nonumber\\
&-&\left.\frac{\lambda_X^-\lambda_X^+}{(\lambda_R^-)^2-(\lambda_X^-)^2}  \left(e^{-2\lambda_X^-(t-t_0)}- e^{-(\lambda_X^-+\lambda_R^-)(t-t_0)}\right)    \right\}.\nonumber
\end{eqnarray}

The major steps of the calculations to obtain $T_2(t-t_0)$ and $T_3(t-t_0)$ are shown in the Appendix.\\

The partial \textbf{contribution} to gene expression noise arising \textbf{from promoter activation} is given by
\begin{eqnarray}
&&\left\langle \left(\langle X(t) \rangle_{X,R} - \langle X(t) \rangle_{X,R,A}\right)^2 \right\rangle_{A}\\ &=& 
\underbrace{\left\langle \langle X(t_0^+) \rangle_{X,R}^2 - \langle X(t_0^+) \rangle_{X,R,A}^2 \right\rangle_{A}e^{-2\lambda_X^-(t-t_0)}}_{J_1}\\
&+& \underbrace{2\int_{t_0^+}^t \left\langle \langle X(t_0^+) \rangle_{X,R} \langle \lambda_X(t')\rangle_{R} - \langle X(t_0^+) \rangle_{X,R,A}\langle \lambda_X(t') \rangle_{R,A} \right\rangle_{A}e^{-\lambda_X^-(t-t_0)}e^{-\lambda_X^-(t-t')}dt'}_{J_2} \label{T2gene} \\
&+& \underbrace{\int_{t_0^+}^t\int_{t_0^+}^t \left\langle \langle \lambda_X(t')\rangle_R \langle \lambda_X(t'')\rangle_R - \langle \lambda_X(t')\rangle_{R,A} \langle \lambda_X(t'') \rangle_{R,A} \right\rangle_{A}e^{-\lambda_X^-(t-t'')}e^{-\lambda_X^-(t-t')}dt'dt''}_{J_3} \label{T3gene}.
\end{eqnarray}

\noindent{\bf Term $J_3$ can be split up into} 
\begin{eqnarray}
J_3(t-t_0)&=&\phi \int_{t_0^+}^t\int_{t_0^+}^{t}\int^{t'}_{-\infty}\int^{t''}_{-\infty}\left[e^{-\gamma_A|\tau'-\tau''|}\left(\frac{1}{2}\right)^{D(t',\tau')}e^{-\lambda_R^-(t'-\tau')}\left(\frac{1}{2}\right)^{D(t'',\tau'')}\right.\nonumber\\
&&\qquad\qquad\quad\times e^{-\lambda_R^-(t''-\tau'')}e^{-\lambda_X^-(t-t')}e^{-\lambda_X^-(t-t'')}\bigg]d\tau''d\tau'dt''dt'\nonumber\\
&=& 2\phi \Big[ \underbrace{\int_{t_0^+}^t\int_{t_0^+}^{t'}\int_{t_0^+}^{t'}\int^{t_0^+}_{-\infty}[.]d\tau''d\tau'dt''dt'}_{I_1}+ \underbrace{ \int_{t_0^+}^t\int_{t_0^+}^{t'}\int^{t_0^+}_{-\infty}\int^{t_0^+}_{-\infty}[.]d\tau''d\tau'dt''dt'}_{I_2}\nonumber\\ &+&\underbrace{\int_{t_0^+}^t\int_{t_0^+}^{t'}\int^{t_0^+}_{-\infty}\int_{t_0^+}^{t''}[.]d\tau''d\tau'dt''dt'}_{I_3} +\underbrace{ \int_{t_0^+}^t\int_{t_0^+}^{t'}\int_{t_0^+}^{t'}\int_{t_0^+}^{t''}[.]d\tau''d\tau'dt''dt'}_{I_4}  \Big]
\end{eqnarray}
\noindent and equivalently {\bf term $J_2$} reads
\begin{eqnarray}
J_2(t-t_0)&=&\phi \int_{t_0^+}^t\int_{-\infty}^{t_0^+}\int^{t'}_{-\infty}\int^{t''}_{-\infty}\left[e^{-\gamma_A|\tau'-\tau''|}\left(\frac{1}{2}\right)^{D(t',\tau')}e^{-\lambda_R^-(t'-\tau')}\left(\frac{1}{2}\right)^{D(t'',\tau'')}\right.\nonumber\\
&&\qquad\qquad\quad\times e^{-\lambda_R^-(t''-\tau'')}e^{-\lambda_X^-(t-t')}e^{-\lambda_X^-(t-t'')}\left(\frac{1}{2}\right)^{D(t_0^+,t'')}\bigg]d\tau''d\tau'dt''dt'
\end{eqnarray}
with
\begin{eqnarray}
\phi&=&var(A)(\lambda_X^+)^2(\lambda_R^+)^2 B_{eq}^2=\frac{var(A)}{\langle A \rangle^2}{\cal A}^2.
\end{eqnarray}
The term $J_2(t-t_0)$ is calculated for the asymptotic cases of interest whereas $J_3(t-t_0)$ can be given by the formulas for $I_1-I_4$.
The explicit expressions to obtain the single terms for $I_1-I_4$ are derived in an equivalent way to the transcriptional contribution (see Appendix).\\

\noindent Calculations lead to the explicit expression for {\bf term $I_1$}, i.e. 
\begin{eqnarray}
I_1&=&\phi_1 \bigg\{ \frac{1}{2\lambda_X^--(\gamma_A + \lambda_R^-)}\left[e^{-(\gamma_A+\lambda_R^-)(t-t_0)}-e^{-2\lambda_X^-(t-t_0)}  \right]\nonumber\\
&-&\frac{1}{\lambda_X^--\gamma_A}\left[ e^{-(\gamma_A+\lambda_X^-)(t-t_0)}-e^{-2\lambda_X^-(t-t_0)} \right]\nonumber\\
&-&\frac{1}{2(\lambda_X^--\lambda_R^-)}\left[ e^{-2\lambda_R^-(t-t_0)}-e^{-2\lambda_X^-(t-t_0)} \right]\nonumber\\
&+&\frac{1}{\lambda_X^--\lambda_R^-}\left[ e^{-(\lambda_R^-+\lambda_X^-)(t-t_0)}-e^{-2\lambda_X^-(t-t_0)} \right]\bigg\} 
\label{T3geneI1}
\end{eqnarray}
with $\phi_1$ given by
\begin{equation}
\phi_1=\frac{1}{2}\frac{1}{(\lambda_R^-)^2-\gamma_A^2}Z_{RA}\frac{1}{\lambda_X^--\lambda_R^-}.
\end{equation}

\noindent{\bf Term $I_2$} is given by
\begin{eqnarray}
I_2&=&\frac{1}{4} \left\{\frac{1}{\lambda_R^-+\gamma_A}\frac{1}{\lambda_R^-}Z_R^{(2)} - \frac{1}{(\lambda_R^-)^2-\gamma_A^2}\frac{e^{-(\lambda_R^-+\gamma_A)T_G}-e^{-2\lambda_R^-T_G}}{(1-2^{-1}e^{-(\gamma_A+\lambda_R^-)T_G})(1-2^{-2}e^{-2\lambda_R^- T_G})}\right\}\nonumber\\
&&\times\frac{1}{2(\lambda_X^--\lambda_R^-)^2}\left[e^{-\lambda_R^-(t-t_0)}-e^{-\lambda_X^-(t-t_0)}\right]^2. 
\end{eqnarray}

\noindent{\bf Term $I_3$} is calculated to be of the form
\begin{eqnarray}
I_3&=&\phi_3 \left\{\frac{1}{\lambda_X^- - \gamma_A}\frac{1}{2\lambda_X^- - (\gamma_A+\lambda_R^-)}\left[e^{-(\lambda_R^-+\gamma_A)(t-t_0)}- e^{-2\lambda_X^-(t-t_0)}\right] \right. \nonumber\\
&&+\left(\frac{1}{\lambda_X^--\lambda_R^-}\left[ \frac{1}{\lambda_X^--\lambda_R^-} - \frac{1}{\lambda_X^- - \gamma_A}\right]\right)\left[e^{-(\lambda_R^- + \lambda_X^-)(t-t_0)}- e^{-2\lambda_X^-(t-t_0)}\right]\nonumber\\
&&- \left.\frac{1}{2(\lambda_X^--\lambda_R^-)^2}\left[e^{-2\lambda_R^-(t-t_0)}- e^{-2\lambda_X^-(t-t_0)}\right]\right\} 
\end{eqnarray}
with
\begin{eqnarray}
\phi_3&=&\frac{1}{2} \frac{1}{(\lambda_R^-)^2-\gamma_A^2}Z_{RA}.
\end{eqnarray}

\noindent The explicit formula for {\bf term $I_4$} is given by 
\begin{eqnarray}
I_4&=&\int_{t_0^+}^t\int_{t_0^+}^{t'}\int_{t_0^+}^{t'}\int_{t_0^+}^{t''}[.] d\tau''d\tau'dt''dt'\nonumber\\
&=&\underbrace{\int_{t_0^+}^t\int_{t_0^+}^{t'}\int_{t_0^+}^{t''}\int_{t_0^+}^{t''}[.]d\tau''d\tau'dt''dt'}_{H_1} + \underbrace{ \int_{t_0^+}^t\int_{t_0^+}^{t'}\int_{t''}^{t'}\int_{t_0^+}^{t''}[.]d\tau''d\tau'dt''dt'}_{H_2}
\end{eqnarray}
\noindent with
\begin{eqnarray}
H_1&=&\frac{2}{\lambda_R^-+\gamma_A}\nonumber\\
&\times&\Bigg\{\frac{1}{2\lambda_R^-(\lambda_X^-+\lambda_R^-)}\left(\frac{1}{2\lambda_X^-}\left[1-e^{-2\lambda_X^-(t-t_0)}\right]- \frac{1}{\lambda_X^--\lambda_R^-}\left[e^{-(\lambda_R^-+\lambda_X^-)(t-t_0)}-e^{-2\lambda_X^-(t-t_0)}\right]\right)\nonumber\\
&+&\left(\frac{1}{2\lambda_R^-(\lambda_X^- - \lambda_R^-)} +\frac{1}{(\lambda_R^--\gamma_A)(\lambda_X^--\gamma_A)} - \frac{1}{\lambda_R^--\gamma_A}\frac{1}{\lambda_X^- - \lambda_R^-}\right)\nonumber\\
&\times&\frac{1}{\lambda_X^- - \lambda_R^-}\left[e^{-(\lambda_R^- + \lambda_X^-)(t-t_0)}-e^{-2\lambda_X^-(t-t_0)} \right]\nonumber\\
&-&\left(\frac{1}{2\lambda_R^-(\lambda_X^- - \lambda_R^-)}-  \frac{1}{\lambda_R^--\gamma_A}\frac{1}{\lambda_X^- - \lambda_R^-} \right)\frac{1}{2(\lambda_X^- - \lambda_R^-)}\left[e^{-2\lambda_R^-(t-t_0)}-e^{-2\lambda_X^-(t-t_0)}\right]\nonumber\\
&-&\frac{1}{(\lambda_R^--\gamma_A)(\lambda_X^--\gamma_A)}\frac{1}{2\lambda_X^- - (\gamma_A + \lambda_R^-)}\left[e^{-(\gamma_A +\lambda_R^-)(t-t_0)}-e^{-2\lambda_X^-(t-t_0)}\right]\Bigg\}
\end{eqnarray}
and
\begin{eqnarray}
H_2&=&\frac{1}{\lambda_R^-+\gamma_A}\frac{1}{\lambda_R^--\gamma_A} \nonumber\\
&\times&\Bigg\{\left(\frac{1}{\lambda_X^-+\gamma_A}- \frac{1}{\lambda_X^- + \lambda_R^-} \right)\frac{1}{2\lambda_X^-}\left[1-e^{-2\lambda_X^-(t-t_0)} \right]\nonumber\\
&+&\left(\frac{1}{\lambda_X^--\lambda_R^-}- \frac{1}{\lambda_X^-+\gamma_A}\right)\frac{1}{\lambda_X^--\gamma_A}\left[e^{-(\gamma_A+\lambda_X^-)(t-t_0)}-e^{-2\lambda_X^-(t-t_0)} \right]\nonumber\\
&-&\left(\frac{1}{\lambda_X^--\lambda_R^-}- \frac{1}{\lambda_X^--\gamma_A}\right)\frac{1}{2\lambda_X^--(\lambda_R^-+\gamma_A)}\left[e^{-(\lambda_R^-+\gamma_A)(t-t_0)}-e^{-2\lambda_X^-(t-t_0)} \right]\nonumber\\
&+&\left(\frac{1}{\lambda_X^-+\lambda_R^-}- \frac{1}{\lambda_X^--\gamma_A}\right)\frac{1}{\lambda_X^--\lambda_R^-}\left[e^{-(\lambda_R^-+\lambda_X^-)(t-t_0)}-e^{-2\lambda_X^-(t-t_0)} \right]\Bigg\}.\label{T3geneH2}
\end{eqnarray}

\section{Asymptotic Analysis of the Partial Contributions}\label{sec3}

\subsection{\bf Case I: Short mRNA lifetime, long protein lifetime $(\lambda_R^-)^{-1}\ll T_G\ll (\lambda_X^-)^{-1}$}
This is the most physiological case and we obtain the asymptotic expressions for the mean molecule number within on cell cycle given in Eqns.~(15), (16) in the main text.
The \textbf{transcriptional contribution} can be split up into the correlations between the actual and previous generations ($T_2$) and the noise from the new generation ($T_3$) and we obtain for the asymptotic case $(\lambda_X^-)^{-1}\gg T_G$
\begin{eqnarray}
T_2(t-t_0)&=&\frac{1}{2}{\cal A}\frac{\lambda_X^+}{(\lambda_R^-)^3}\left(1-e^{-\lambda_R^-(t-t_0)} \right). 
\end{eqnarray}
and
\begin{eqnarray}
T_3(t-t_0)= 2 {\cal A}\frac{\lambda_X^+}{(\lambda_R^-)^2} \left[(t-t_0)-\frac{3}{2}\frac{1}{\lambda_R^-}\left(1-e^{-\lambda_R^-(t-t_0)} \right) \right].
\end{eqnarray}
Including the contribution from $T_1$ and assuming $t-t_0 \gg (\lambda_R^-)^{-1}$, we obtain for the transcriptional contribution to gene expression noise the expression of Eqn.~(19) in the main text

The single expression for the \textbf{contributions from promoter activation} are given by 

\begin{eqnarray}
J_2(t-t_0) &=& {\cal A}^2\frac{var(A)}{\langle A\rangle^2} \frac{1}{(\lambda_R^-)^2-(\gamma_A)^2}\left(\frac{1}{\gamma_A^2}Z_{A}^{(1)}\left[1-e^{-\gamma_A(t-t_0)}\right]\right.\nonumber\\
&& \qquad\qquad\qquad \left.- \frac{\gamma_A}{(\lambda_R^-)^3}\left[1-e^{-\lambda_R^-(t-t_0)}\right]\right) \label{GA2}
\end{eqnarray}
and
\begin{eqnarray}
J_3(t-t_0)&=&2\phi H_2 =2 {\cal A}^2\frac{var(A)}{\langle A\rangle^2}\frac{1}{(\lambda_R^-)^2-(\gamma_A)^2}\left[\frac{1}{\gamma_A}(t-t_0) - \frac{1}{\gamma_A^2}(1-e^{-\gamma_A(t-t_0)})\right].
\label{GA3}
\end{eqnarray}

Assuming $t-t_0 \gg (\gamma_A)^{-1}, (\lambda_R^-)^{-1}$ and including the contributions from $J_1$, we obtain the noise contribution from promoter activation given in Eqn.~(20) in the main text.

\subsection{\bf Case II: Long mRNA and protein lifetime $(\lambda_R^-)^{-1},(\lambda_X^-)^{-1}\gg T_G$}
 
The terms for the \textbf{transcriptional contribution} read
\begin{eqnarray}
T_2(t-t_0)&=&\frac{1}{3}{\cal A}\lambda_X^+ T_G^2 (t-t_0).
\end{eqnarray}
and noise from the new generation is given by
\begin{eqnarray}
T_3(t-t_0)&=& \frac{1}{3}{\cal A}\lambda_X^+ (t-t_0)^3.
\end{eqnarray}


The \textbf{contributions from promoter activation} read
\begin{eqnarray}
J_2(t-t_0)&=&{\cal A}^2 \frac{var(A)}{\langle A\rangle^2}\frac{Z_A^{(1)}}{\gamma_A}\left\{8T_G^2(t-t_0)+\frac{4}{3\gamma_A}T_G(t-t_0)\right.\nonumber\\
&&\qquad\qquad\qquad\quad\left. -\frac{1}{\gamma_A^2}(1-e^{-\gamma_A(t-t_0)})\frac{1-e^{-\gamma_AT_G}}{1-\frac{1}{4}e^{-\gamma_AT_G}}\right\}\label{Case2J2promact}
\end{eqnarray}
and
\begin{eqnarray}
J_3(t-t_0)&=&2{\cal A}^2 \frac{var(A)}{\langle A\rangle^2}(I_1+I_2+I_3+I_4)
\end{eqnarray}
with
\begin{eqnarray}
I_1&=&\frac{1}{4\gamma_A^2}Z_A^{(1)}(t-t_0)^2 \\
I_2&=&\frac{1}{6\gamma_A}\left[2T_G- \frac{Z_A^{(1)}}{\gamma_A}\right](t-t_0)^2\\
I_3&=&\frac{Z_A^{(1)}}{2\gamma_A^2}\left\{\frac{1}{\gamma_A^2}\left(1-e^{-\gamma_A(t-t_0)}\right) -\frac{t-t_0}{\gamma_A}+ \frac{(t-t_0)^2}{2}\right\} \\
I_4&=&\frac{(t-t_0)^3}{3\gamma_A}-\frac{(t-t_0)^2}{2\gamma_A^2}.
\end{eqnarray}

In the asymptotic limit (for $T_G,(t-t_0)\gg \gamma_A^{-1}$) we obtain Eqns.~(23) and (24) in the main text for the noise due to transcription and promoter activation, respectively.

\subsection{\bf Case III: Short mRNA and short protein lifetime $(\lambda_R^-)^{-1},(\lambda_X^-)^{-1}\ll T_G$}
The mean amounts of protein and mRNA are given by Eqns.~(25) and (26) in the main text. 
The terms for the \textbf{transcriptional contribution} are given in the limit $T_G \gg (\lambda_R^-)^{-1},(\lambda_X^-)^{-1}$ by
\begin{eqnarray}
T_2(t-t_0)&=&\frac{2}{3}{\cal A} \frac{\lambda_X^+}{\lambda_R^-((\lambda_R^-)^2-(\lambda_X^-)^2)} \left(e^{-2\lambda_X^-(t-t_0)}-e^{-(\lambda_R^-+\lambda_X^-)(t-t_0)}\right).
\end{eqnarray}
and
\begin{eqnarray}
T_3(t-t_0)&=& Q_X B_{eq} \left\{\left[-\frac{\lambda_X^+}{\lambda_R^--2\lambda_X^-}\left(e^{-2\lambda_X^-(t-t_0)}-\ e^{-\lambda_R^-(t-t_0)}\right) \right.\right.\nonumber\\
 &+&\left.\left. \frac{\lambda_X^+}{\lambda_R^--\lambda_X^-}\left(e^{-2\lambda_X^-(t-t_0)}- e^{-(\lambda_X^-+\lambda_R^-)(t-t_0)}\right)\right] + \frac{\lambda_X^+}{2(\lambda_R^-+\lambda_X^-)}\left(1-e^{-2\lambda_X^-(t-t_0)}\right)\right.\nonumber\\
&-&\left.\frac{\lambda_X^-\lambda_X^+}{(\lambda_R^-)^2-(\lambda_X^-)^2}  \left(e^{-2\lambda_X^-(t-t_0)}- e^{-(\lambda_X^-+\lambda_R^-)(t-t_0)}\right)    \right\}.
\end{eqnarray}

The only \textbf{contribution from promoter activation} derives from the $I_4$ term ($I_1=I_2=I_3=0$).
As the fluctuations decay very fast we also have no contribution from previous generations ($J_1=0$)
and no correlations between the previous and actual generation ($J_2=0$).\\

We obtain
\begin{eqnarray}\label{sigmaAcase4}
J_3(t-t_0)&=& 2{\cal A}^2 \frac{var(A)}{\langle A\rangle^2}I_4
=\frac{{\cal A}^2 var(A)}{\langle A\rangle^2\lambda_R^-\lambda_X^-(\gamma_A+\lambda_R^-)(\lambda_R^-+\lambda_X^-)}\left(1+\frac{\lambda_R^-}{\gamma_A+\lambda_X^-}\right).
\end{eqnarray}
Eqns.~(27) and (28) of the main text are obtained in the limit $t-t_0\gg (\gamma_A)^{-1}$.

\section{Simulating partial noise contributions}\label{sec4}

We investigate the partial contributions (terms of Eqn.~(\ref{noisecont})) to gene expression noise by the following simulation protocol.
If we assume a constant activator $B_{eq}=const$, there are no contributions from activator binding.
In the more general case, i.e. if $B(t)\neq const$, it can be easily included in the simulations.\\
As mentioned before, we can also neglect the noise contribution from translation.
The remaining terms (transcription) and (promoter activation) only differ in their averaging.

In order to calculate the partial contributions to gene expression noise, a correct averaging of the simulation results is necessary:\\
Let $g,m$ be the number of simulations of promoter-\textit{on/-off} trajectories and mRNA trajectories, respectively.
Each protein trajectory has a unique promoter-\textit{on/-off} and mRNA trajectory, so $p_{ij}$ is the protein trajectory resulting from the $i$th promoter-\textit{on/-off} and the $j$th mRNA trajectory and can be identified with $\langle X(t)\rangle_X$.\

Averaging over all mRNA trajectories belonging to one promoter-\textit{on/-off} trajectory, we obtain the mean protein trajectory belonging to the $i$th promoter-\textit{on/-off} realization, i.e.
$ \langle X(t)\rangle_{X,R}=\frac{1}{m}\sum_{j=1}^{m}p_{ij}$.\

Averaging over all mRNA trajectories which were obtained from all promoter-\textit{on/-off} trajectories we have
$\langle X(t)\rangle_{X,R,A}=\frac{1}{g}\sum_{i=1}^{g}\frac{1}{m}\sum_{j=1}^mp_{ij}.$\

We obtain for the second moment of transcription ($\sigma_R^2$) and promoter activation ($\sigma_A^2$) of gene expression noise on the protein level:
\begin{eqnarray}
\sigma_R^2&=&\left\langle \left(\langle X(t)\rangle_{X}-\langle X(t)\rangle_{X,R}\right)^2\right\rangle_{R,A,B}=\left\langle \left(\langle X(t)\rangle_{X}-\langle X(t)\rangle_{X,R}\right)^2\right\rangle_{R,A}\nonumber\\
&=&\frac{1}{g}\sum_{i=1}^{g}\frac{1}{m}\sum_{j=1}^{m}(p_{ij}- \frac{1}{m}\sum_{j=1}^{m}p_{ij})^2 \\
\sigma_A^2&=&\left\langle \left(\langle X(t)\rangle_{X,R}-\langle X(t)\rangle_{X,R,A}\right)^2\right\rangle_{A,B} 
=\left\langle \left(\langle X(t)\rangle_{X,R}-\langle X(t)\rangle_{X,R,A}\right)^2\right\rangle_{A} \nonumber\\
&=&\frac{1}{g}\sum_{i=1}^{g}(\frac{1}{m}\sum_{j=1}^{m}p_{ij} - \frac{1}{g}\sum_{i=1}^{g} \frac{1}{m}\sum_{j=1}^{m}p_{ij})^2.
\end{eqnarray}

To illustrate that this averaging method mimics the correct formulas, if mRNA and protein degradations are large compared to the generation time (\textit{case III}), i.e. the system equilibrate very fast after cell division, Table~\ref{table1} summarizes the mean amount of mRNA and protein and the noise contributions from transcription and promoter activation (averaged over the last half of the cell cycle).
If, however, one of these two rates is small compared to the generation time (\textit{case I} and \textit{II}), the simulation/averaging routine becomes much more complex and very time consuming.
To assure that the analytical expressions are correct, we calculated some values of the integral representation of the partial contributions (Eqns. \eqref{T2gene}, \eqref{T3gene}) with Mathematica and compared them to the explicit formulas (Eqns. (\ref{GA2}),\eqref{Case2J2promact}) and (\ref{T3geneI1})-(\ref{T3geneH2})).
For \textit{case I} and \textit{II}, the precise formulas evaluated at three different time points coincide with the integral representations at these time points.

\section{Appendix}

We derive the explicit formulas for the transcriptional contribution, i.e. Eqns. (\ref{T2trans},\ref{T3trans}) and give some calculations leading to (\ref{T2transana}) and (\ref{T3transana}).
The contribution from promoter activation is obtained in an equivalent way.
\begin{eqnarray*}
T_2(t-t_0)&=&(\lambda_X^+)^2\int_{t_0}^t\int_{-\infty}^{t_0^-}\big\langle\underbrace{\langle R(t'')R(t')\rangle_R-\langle R(t'')\rangle_R\langle R(t')\rangle_R\rangle}_{\langle R(t'')\rangle_Re^{-\lambda_R^-(t'-t'')}2^{-D(t',t'')}}\big\rangle_A\\
&&\qquad\qquad\qquad \times e^{-\lambda_X^-(2t-t'-t'')}2^{-D(t_0^-,t'')} dt''dt'\\
&=& \frac{(\lambda_X^+)^2\lambda_R^+}{\lambda_R^-}\langle A\rangle B_{eq} \int_{t_0^-}^t \int_{-\infty}^{t_0^-}\Big[1-e^{-\lambda_R^-(t''-t_0)}(1-2^{-1}Z_R^{(1)})\Big] \\ 
&\times&e^{-(\lambda_X^-+\lambda_R^-)(t'-t'')}e^{-2\lambda_X^-(t-t')}\frac{1}{2}4^{-D(t_0^-,t'')}dt''dt' \\
&=& \frac{(\lambda_X^+)^2\lambda_R^+}{2\lambda_R^-} \langle A\rangle B_{eq} \int_{t_0^-}^t \sum_{k=0}^{\infty} 4^{-k} \int_{t_0-(k+1)T_G}^{t_0-kT_G}\Big[1-e^{-\lambda_R^-(t''-(t_0-(k+1)T_G))}(1-2^{-1}Z_R^{(1)})\Big]\\ 
 &\times&e^{-(\lambda_X^-+\lambda_R^-)(t'-t'')}e^{-2\lambda_X^-(t-t')}dt''dt'\\
&=& \frac{(\lambda_X^+)^2\lambda_R^+}{2\lambda_R^-} \langle A\rangle B_{eq} \int_{t_0^-}^t\left[\frac{1-e^{-(\lambda_X^-+\lambda_R^-)G}}{(\lambda_X^-+\lambda_R^-)(1-\frac{1}{4}e^{-(\lambda_X^-+\lambda_R^-)T_G})}\right. \\
&-&\left.\frac{e^{-\lambda_R^-T_G}-e^{-(\lambda_X^-+\lambda_R^-)T_G}}{\lambda_X^-(1-\frac{1}{4}e^{-(\lambda_X^-+\lambda_R^-)T_G})} (1-2^{-1}Z_R^{(1)})\right]*e^{-(\lambda_X^-+\lambda_R^-)t'}e^{2\lambda_X^-(t-t')}dt' \\
&=& (\ref{T2transana})
\end{eqnarray*}
\begin{eqnarray*}
T_3(t-t_0)&=&2(\lambda_X^+)^2\int_{t_0}^t\int_{t_0}^{t'}\langle R(t'')\rangle_{R,A}e^{-\lambda_R^-(t'-t'')}e^{-\lambda_X^-(2t-t'-t'')}dt''dt' \\
&=& \frac{2(\lambda_X^+)^2\lambda_R^+}{\lambda_R^-}\langle A\rangle B_{eq}\int_{t_0}^t\int_{t_0}^{t'} \big[(2^{-1}Z_R^{(1)}-1)e^{-\lambda_R^-(t''-t_0)}+1\big]e^{-\lambda_R^-(t'-t'')}\\
&&\qquad\qquad\qquad\times e^{-\lambda_X^-(2t-t'-t'')}dt''dt' \nonumber \\
&=& 2\lambda_X^+Q_X B_{eq} \int_{t_0}^t\Big[(2^{-1}Z_R^{(1)}-1)e^{-\lambda_R^-(t'-t_0)}(1-e^{-\lambda_X^-(t'-t_0)})e^{-2\lambda_X^-(t-t')} \nonumber \\
&+&\frac{\lambda_X^-}{\lambda_R^-+\lambda_X^-}(1-e^{-(\lambda_X^-+\lambda_R^-)(t'-t_0)})e^{-2\lambda_X^-(t-t')}\Big]dt' \nonumber\\
&=&2\lambda_X^+Q_X B_{eq}\Big\{ \frac{1}{2(\lambda_X^-+\lambda_R^-)}+\frac{1-2^{-1}Z_R^{(1)}}{\lambda_R^--2\lambda_X^-}e^{-\lambda_R^-(t-t_0)} \nonumber\\
&+& e^{-2\lambda_R^-(t-t_0)}\left[(1-2^{-1}Z_R^{(1)})\left( \frac{1}{\lambda_R^--2\lambda_X^-}-\frac{1}{\lambda_X^-+\lambda_R^-}\right)-\frac{1}{2(\lambda_X^-+\lambda_R^-)}\right.\\
&-&\left. \frac{\lambda_X^-}{(\lambda_X^-)^2-(\lambda_R^-)^2}\right] + e^{-(\lambda_R^-+\lambda_X^-)(t-t_0)}\left( 2^{-1}Z_R^{(1)}-1\right) \frac{1}{\lambda_X^-+\lambda_R^-}+\frac{\lambda_X^-}{(\lambda_X^-)^2-(\lambda_R^-)^2}\Big\} \nonumber \\
&=& (\ref{T3transana})
\end{eqnarray*}

\begin{table}
\begin{tabular}{c|c|c}
& analytics & simulations \\ \hline\hline
$\langle R(t)\rangle_{\bf I}$ & $5.33$ & $5.31\pm0.03$ \\
$\langle X(t)\rangle_{\bf I}$ & $106.67$ & $106.37\pm0.39$\\
$\eta_R^2$ & $0.1071$ & $0.1074\pm4.6\times 10^{-4}$\\
$\eta_A^2$ & $0.2402$ & $0.2443\pm2.7\times 10^{-3}$\\ \hline
\end{tabular}
\caption{\label{table1} Comparison of the analytical and simulation results of \textit{case III} with the kinetic parameters $\lambda_A^+=0.2, \lambda_A^-=0.3, \lambda_R^+=2, \lambda_R^-=0.15, \lambda_X^+=4$, $\lambda_X^-=0.2$ and $B_{eq}=1$. The cells divide after 100 minutes.\vspace*{1cm}}
\end{table}

\newpage

\begin{figure}
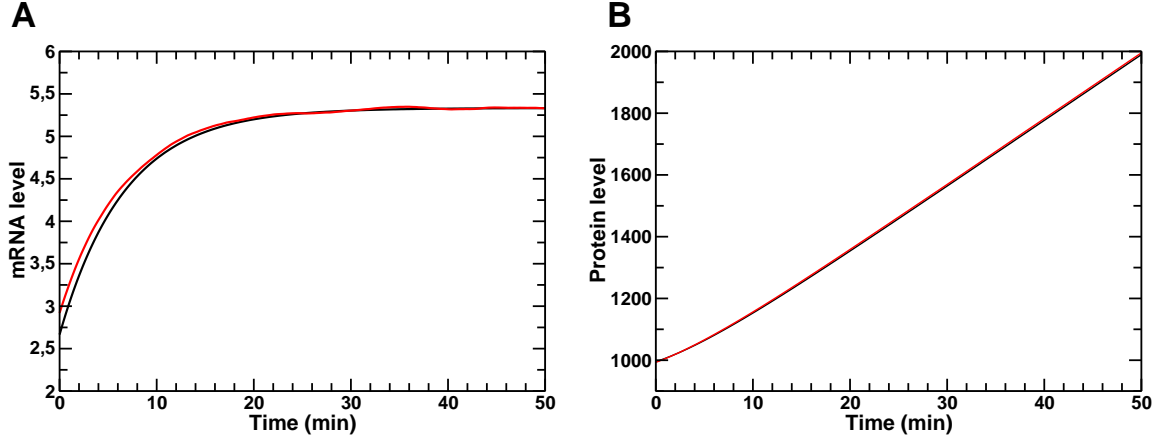

\vspace*{-0mm}\hspace*{3mm}\includegraphics[width=73mm]{case1mRNA.eps}
\vspace*{-0mm}\hspace*{3mm}\includegraphics[width=75mm]{case1protein.eps}
\caption{\label{compareanasimu}Comparison of analytical results (black lines) with simulations (red lines). (A) mean amount of mRNA and (B) mean amount of protein.
The kinetic parameters are chosen from \textit{case I}, when mRNA mRNA lifetime is short and protein lifetime is long compared to the generation time, i.e. $\lambda_A^+=0.2, \lambda_A^-=0.3, \lambda_R^+=2, \lambda_R^-=0.15, \lambda_X^+=4$ and $\lambda_X^-=1\times 10^{-5}$.
The activator is constantly bound, i.e. $B_{eq}=1$, but the same results are also obtained if $B_{eq}=q<1$ (simulations not shown).
We fix the cell cycle to $t_0=0$, $T=50$ (min) and sampling frequency to $dt=0.2$.
To obtain a good agreement between simulations and analytics, we simulate and average over 100 cells, each with 200 promoter-on/-off trajectories. 
For each promoter-on/-off trajectory we generate 200 mRNA trajectories with the corresponding protein trajectories.}
\end{figure}